\documentclass[twocolumn,showpacs,preprintnumbers,amsmath,amssymb,
prd,a4paper,floatfix]{revtex4}
\usepackage{epsfig,hyperref}
\newcommand{\rb}[1]{\raisebox{1.5ex}[-1.5ex]{#1}}
\newcommand\T{\rule{0pt}{2.6ex}}
\newcommand\B{\rule[-1.2ex]{0pt}{0pt}}

\begin{document}

\title{Derivative sources in lattice spectroscopy\\ of excited light-quark mesons}

\author{Christof Gattringer}
\email{christof.gattringer@uni-graz.at}
\affiliation{Institut f\"ur Physik, Unversit\"at Graz, 8010 Graz, Austria}
\author{Leonid Ya. Glozman}
\email{leonid.glozman@uni-graz.at}
\affiliation{Institut f\"ur Physik, Unversit\"at Graz, 8010 Graz, Austria}
\author{C. B.~Lang}
\email{christian.lang@uni-graz.at}
\affiliation{Institut f\"ur Physik, Unversit\"at Graz, 8010 Graz, Austria}
\author{Daniel Mohler}
\email{daniel.mohler@uni-graz.at}
\affiliation{Institut f\"ur Physik, Unversit\"at Graz, 8010 Graz, Austria}
\author{Sasa Prelovsek}
\email{sasa.prelovsek@ijs.si}
\affiliation{Department of Physics, University of Ljubljana and Institut
  Joszef Stefan, 1001 Ljubljana, Slovenia}

\date{\today}

\begin{abstract}
We construct efficient interpolating fields for lattice spectroscopy of mesons 
by applying covariant derivatives on Jacobi smeared quark sources. These
interpolators are tested in a quenched calculation of excited 
mesons based on the variational method. We present results for pseudoscalar,
scalar, vector and pseudovector mesons.
\end{abstract}


\maketitle

\section{Introduction}

A clean extraction of excited hadron masses from a lattice QCD
simulation is a serious challenge. However, an ab-initio
determination of properties for excited light hadrons would provide highly
interesting information on the chiral dynamics of QCD. 

Excited hadrons are rather non-trivial objects to study on the
lattice. One of the main reasons for difficulties is the fact that excited
states appear only as sub-leading contributions in Euclidean two-point
functions. Although a variety of other approaches has been
tried, including Bayesian methods 
\cite{Asakawa:2000tr,Lepage:2001ym,Chen:2004gp}, 
an NMR-inspired blackbox method \cite{Fleming:2004hs,Lin:2007iq} and
evolutionary fitting techniques \cite{vonHippel:2007dz}, the most powerful
method is probably the variational approach
\cite{Michael:1985ne,Luscher:1990ck}. The reason for its power is the fact
that in the variational method not only a single correlator is studied, but a
whole matrix of correlation functions. Consequently more information is
extracted from the system. 

The successful implementation of the variational method hinges crucially on the set of basis
interpolators that are used in the correlation matrix. A particularly
important criterion is that the interpolators have a large overlap with the
physical states in a given channel, i.e., with both, ground- and excited
states. In this article we build on earlier work \cite{Burch:2006dg,
Burch:2006cc, Burch:2005vn, Burch:2004he}, where Jacobi-smeared quark sources
with different width were used to construct hadron interpolators that allow for nodes in
their radial wave function. To construct a richer set of interpolators, we now
also include derivative sources for light-quark spectroscopy.  

Interpolators with derivatives have been widely used for heavy quark systems
(see for example \cite{Liao:2002rj,Dudek:2007wv,Wingate:1995hy}). They have also been applied
by Burch et al.~\cite{Burch:2006rb,Burch:2007dh} for light mesons and by Lacock et
al.~\cite{Lacock:1996vy} to the study of 
both orbital excitations of mesons and hybrids. Note that the
approach in \cite{Lacock:1996vy} differs from ours. There interpolators
are built 
with quarks displaced relative to each other connected by certain paths which are
classified with respect to irreducible representations of the symmetry group
of the hypercubic lattice. That approach is similar to the one adopted by
Basak et al.~\cite{Basak:2005ir,Basak:2005aq,Basak:2007kj,Basak:2006ww} for baryons.

In our paper we test derivative sources in a quenched excited 
meson spectroscopy calculation. In particular we study the pseudoscalar, scalar,
vector and pseudovector channels. Depending on the channel we find
considerable improvement of the signal for some of the excited and ground
states. Preliminary results with our derivative sources were already 
reported in \cite{Gattringer:2007td,Gattringer:2007da,Lang:2007mq}.

\section{Setting of the calculation}

\subsection{Variational method}
The central idea of the variational method
\cite{Michael:1985ne,Luscher:1990ck} is to use several different interpolators
$O_i, i = 1, \ldots \, N$ 
with the quantum numbers of the desired state and to compute all cross
correlators for interpolators projected to fixed spatial momentum (zero in
this work),
\begin{equation}
C(t)_{ij} = \langle O_i(t) O^\dagger_j(0) \rangle . 
\label{corrmatdef}
\end{equation}
In Hilbert space these correlators have the decomposition 
\begin{equation}
C(t)_{ij} = \sum_n \langle 0 | O_i | n \rangle 
\langle n | O_j^\dagger | 0 \rangle e^{-t M_n} . 
\label{corrmatrix}
\end{equation}
Using the factorization of the amplitudes one can show \cite{Luscher:1990ck} 
that the eigenvalues 
$\lambda_k(t)$ of the generalized eigenvalue problem 
\begin{equation}
C(t) \vec{v}_k = \lambda_k(t) C(t_0) 
\vec{v}_k , 
\label{generalized}
\end{equation}
behave as 
\begin{equation}
\lambda_k(t) \propto e^{-t \, M_k}[ 1 +
{\cal O}(e^{-t \Delta M_k})], 
\label{eigenvaluedecay}
\end{equation}
where $M_k$ is the mass of the $k$-th state and $\Delta M_k$ is the difference 
to neighboring states. In Eq.\ (\ref{generalized}) 
the eigenvalue problem is normalized at a timeslice $t_0 \le t$.

Equation (\ref{eigenvaluedecay}) shows that each eigenvalue predominantly 
decays with a single mass: The largest eigenvalue decays with the mass of the 
ground state, the second largest eigenvalue with the mass of the first 
excited state, and so on. Thus, the variational method disentangles the  
signals of the ground- and excited states. As a consequence simple, 
stable two-parameter fits become possible.

At this point we remark, that the variational method also treats ghost
contributions correctly, which in some channels show up in a 
quenched or partially quenched calculation at small quark masses 
\cite{Bardeen:2001jm}. It was shown in \cite{Burch:2005wd} that 
in the variational approach the ghost contribution couples to an individual 
eigenvalue (up to the correction term) in the same way as a proper
physical state. Thus, ghost contributions are disentangled from the 
physical states and need not be modeled in the further analysis of the
exponential decay of the eigenvalues.

To visualize the results and to determine possible fit ranges, we plot
effective masses which are built from the ratios of eigenvalues
\begin{align}
a\,M_{k,\,\mathrm{eff}}\left(t+\frac{1}{2}\right) &=
\mathrm{ln}\left(\frac{\lambda_k(t)}{\lambda_k(t+1)}\right)\ .
\end{align}
For those values of $t$ where the exponential decay of the eigenvalue is
governed by a single state, the effective masses form pronounced plateaus. 

It is an interesting observation that for the same values of $t$, where the
effective mass plateaus form, also the corresponding eigenvectors are
approximately constant as a function of $t$. An example of this behavior is
given in Fig.~\ref{fig_pion_eigenvectors} (discussed later), where we show the entries of the eigenvectors for the three largest
eigenvalues as a function of $t$. This time-independence of the eigenvectors
serves as a ``fingerprint'' for the physical states. To be more
precise, the eigenvectors we use for such fingerprints are the eigenvectors
of the regular eigenvalue problem
\begin{align}
C(t_0)^{-\frac{1}{2}}C(t)C(t_0)^{-\frac{1}{2}}\vec{v}^{\,\prime}_k&=
\lambda_k(t)\vec{v}^{\,\prime}_k\ ,
\label{normalev}
\end{align}
which obviously has the same eigenvalues $\lambda_k(t)$ 
as the problem (\ref{generalized}),
but gives rise to orthogonal eigenvectors $\vec{v}^{\,\prime}_k$.

In our analysis we only fit states which give rise to 
a plateau in both the effective mass and the
corresponding eigenvector. Ideally, $t_0$ should be chosen large. However,
large $t_0$ also tends to increase the statistical noise.
We explored the dependence of the effective mass
plateaus on the timeslice $t_0$ and usually found the best results for
$t_0=1$. So unless noted otherwise, $t_0$ will be fixed to $t_0 = 1$ 
for all results quoted (our sources are located at $t=0$; see next section).

\subsection{Smeared sources and sinks}

In order to optimize the overlap with the ground and first few excited states, one
commonly uses quark smearing. We first construct extended sources by Jacobi
smearing \cite{Gusken:1989ad, Best:1997qp} of point sources 
$S_0$ located at timeslice $t=0$:
\begin{align}
S_0^{(\alpha,a)}(\vec{y},t)_{\rho,c}&=\delta(\vec{y},\vec{0}\,) 
\delta(t,0)\delta_{\rho
  \alpha}\delta_{ca}\ ,\\
S^{(\alpha,a)}&=\sum_{n=0}^N\kappa^nH^n S_0^{(\alpha,a)} \ ,\\
H(\vec{x},\vec{y})&=\sum_{i=1}^3\left(U_i(\vec{x},0)
\delta(\vec{x}+\hat{i},\vec{y}\,)\right.\nonumber\\
&\qquad\left.+U_i(\vec{x}-\hat{i},0)^\dagger\delta(\vec{x}-\hat{i},
\vec{y}\,)\right)\ .
\end{align}
The smearing has two parameters $\kappa$ and $N$ and leads to
gauge covariant, 
approximately Gaussian shaped sources of different width. We use the same
combinations of parameters as in \cite{Burch:2006dg} and we refer to our
sources as ``narrow'' ($S_n$) and ``wide'' ($S_w$). 

Our derivative quark sources $W_{\partial_i}$ are constructed 
by applying a covariant derivative to the wide
sources:
\begin{align}
P_i(\vec{x},\vec{y})&=U_i(\vec{x},0)\delta(\vec{x}+\hat{i},\vec{y}\,)-
U_i(\vec{x}-\hat{i},0)^\dagger\delta(\vec{x}-\hat{i},\vec{y}\,)\
,\\
W_{\partial_i}&=P_iS_w\ .
\end{align}
These sources are then used in the construction of meson interpolators of
definite quantum numbers.

\begin{figure*}[bt]
\includegraphics[height=6.7cm,clip]{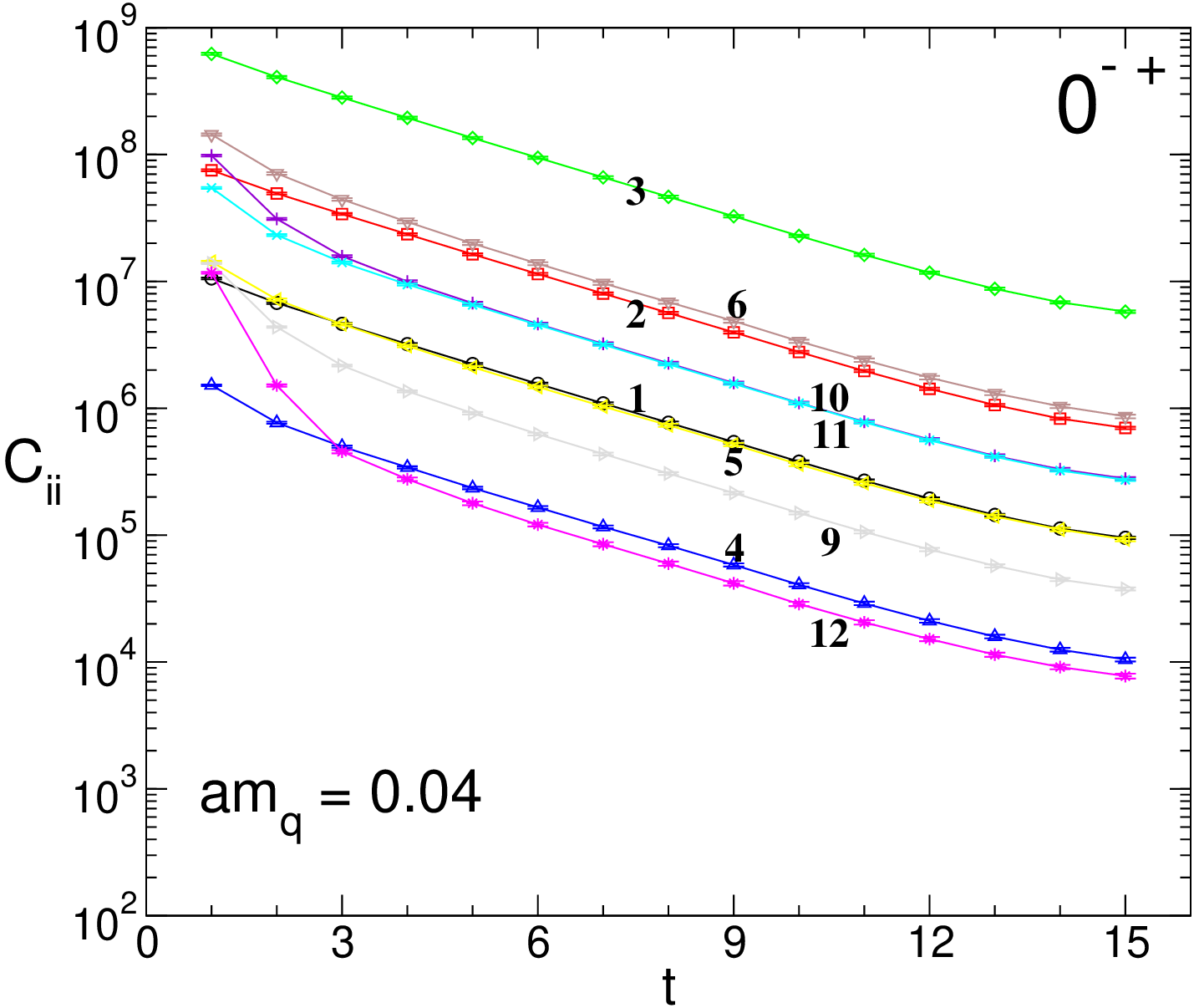} \hspace{10mm}
\includegraphics[height=6.7cm,clip]{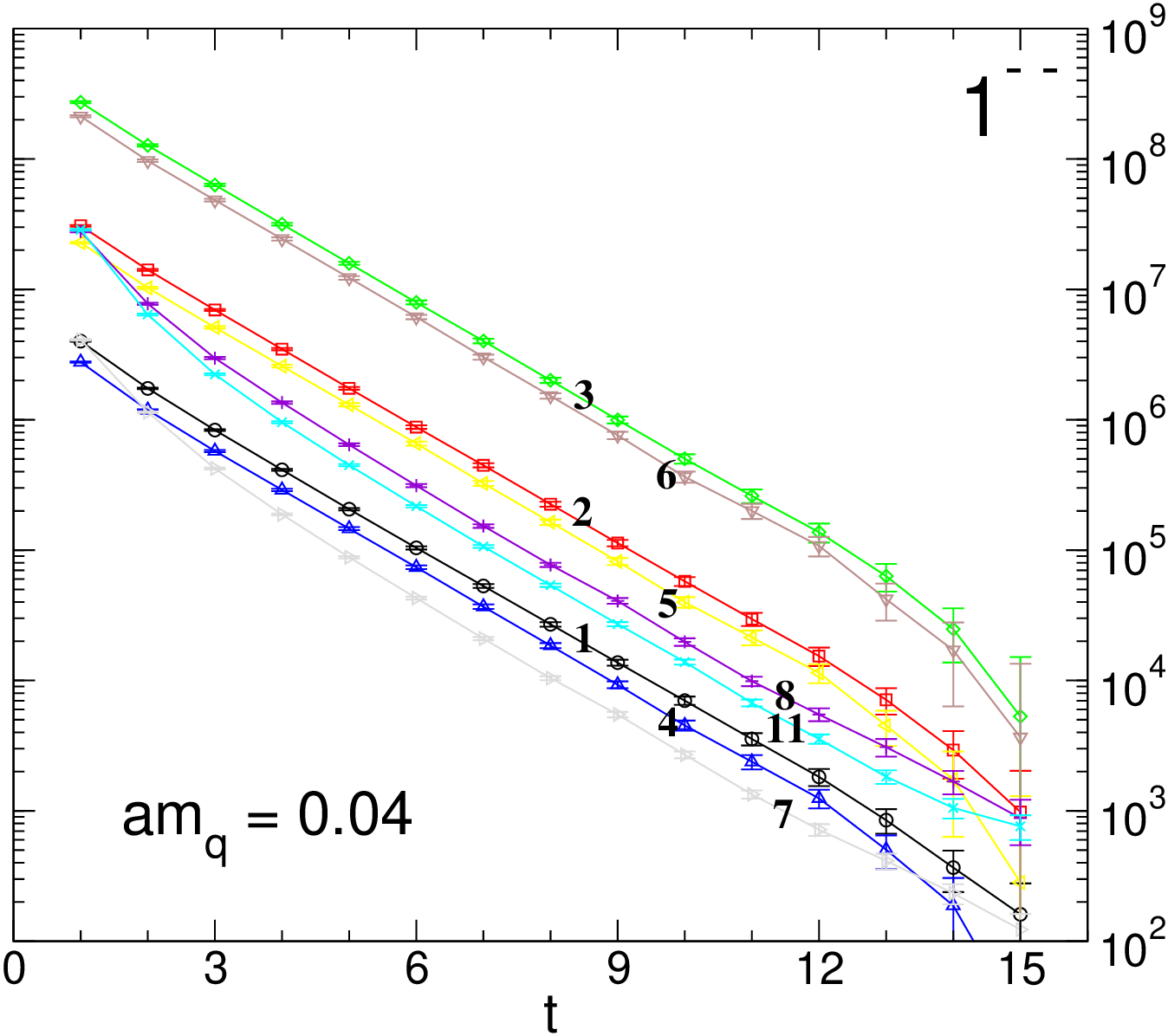}
\caption{
Diagonal entries of the correlation matrix as a function of $t$. The data are for 
bare quark mass $am_q = 0.04$. The $0^{-+}$ (lhs.~plot) and $1^{--}$ (rhs.)
channels are shown. The numbers next to the correlators are according to Table \ref{interpolatortable}.
}
\label{fig_distribution}
\end{figure*}

\begin{table*}[tb]
\begin{center}
\begin{ruledtabular}
\begin{tabular}{c|c|c|c|c|p{6.3cm}}
Nr. & operator & $I\;J^{PC}$ & chiral representation & $I\;J^{PC}$ &
comment \\
& & beyond chiral limit &  & chiral limit & \\
\hline\hline
\T\B 1 & $\overline{u}_n d_n$ & & & & \\
\cline{1-2}
\T\B 2 & $\overline{u}_n d_w$ & $1\;0^{++}$ & $\left(\frac{1}{2},\frac{1}{2}\right)_b$ &
$1\;0^{++}$ & \\
\cline{1-2}
\T\B 3 & $\overline{u}_w d_w$ & & & & \\
\hline
\T\B 7 & $\overline{u}_{\partial_i} \gamma_i d_n$ & & & \raisebox{-0.5ex}[-1.5ex]{does not exist} &\\
\cline{1-2}
\T\B 8 & $\overline{u}_{\partial_i} \gamma_i d_w$ & \rb{$1\;0^{++}$} &
\rb{$(1,0)\oplus (0,1)$} & \raisebox{0.5ex}[-1.5ex]{(only $J\ge 1$)} & \rb{not
  coupling to scalar in the chiral limit}\\
\hline
\T\B 9 & $\overline{u}_{\partial_i} \gamma_i \gamma_4 d_n$ & & & &\\
\cline{1-2}
\T\B 10 & $\overline{u}_{\partial_i} \gamma_i \gamma_4 d_w$ &  \rb{$1\;0^{++}$} &
\rb{$\left(\frac{1}{2},\frac{1}{2}\right)_b$} & \rb{$1\;0^{++}$} & \\
\hline
\T\B 11 & $\overline{u}_{\partial_i} d_{\partial_i}$ & $1\;0^{++}$ &
$\left(\frac{1}{2},\frac{1}{2}\right)_b$ & $1\;0^{++}$ & \\
\hline\hline
\T\B 1 & $\overline{u}_n \gamma_5 d_n$ & & & &\\
\cline{1-2}
\T\B 2 & $\overline{u}_n \gamma_5 d_w$ & $1\;0^{-+}$ &
$\left(\frac{1}{2},\frac{1}{2}\right)_a$ & $1\;0^{-+}$ & \\
\cline{1-2}
\T\B 3 & $\overline{u}_w \gamma_5 d_w$ & & & &\\
\hline
\T\B 4 & $\overline{u}_n \gamma_4 \gamma_5 d_n$ & & & &\\
\cline{1-2}
\T\B 5 & $\overline{u}_n \gamma_4 \gamma_5 d_w$ & $1\;0^{-+}$ & $(1,0)\oplus (0,1)$
& \raisebox{1.0ex}[-1.5ex]{does not exist} & \raisebox{1.0ex}[-1.5ex]{time component of axial vector coupling} \\
\cline{1-2}
\T\B 6 & $\overline{u}_w \gamma_4 \gamma_5 d_w$ & & &
\raisebox{2.0ex}[-1.5ex]{(only $J\ge 1$)} & \raisebox{2.0ex}[-1.5ex]{due to chiral symmetry
breaking}\\
\hline
\T\B 9 & $\overline{u}_{\partial_i} \gamma_i \gamma_4 \gamma_5 d_n$ & & & & \\
\cline{1-2}
\T\B 10 & $\overline{u}_{\partial_i} \gamma_i \gamma_4 \gamma_5 d_w$ &
\rb{$1\;0^{-+}$} & \rb{$\left(\frac{1}{2},\frac{1}{2}\right)_a$} &
\rb{$1\;0^{-+}$} & \\
\hline
\T\B 11 & $\overline{u}_{\partial_i} \gamma_5 d_{\partial_i}$ & $1\;0^{-+}$ &
$\left(\frac{1}{2},\frac{1}{2}\right)_a$ & $1\;0^{-+}$ & \\
\hline
\T\B 12 & $\overline{u}_{\partial_i} \gamma_4\gamma_5 d_{\partial_i}$ & $1\;0^{-+}$
& $(1,0)\oplus (0,1)$ & does not exist & time component of axial
vector coupling \\
& & & & \raisebox{1.0ex}[-1.5ex]{(only $J\ge 1$)} & \raisebox{1.0ex}[-1.5ex]{due to chiral symmetry breaking}\\
\hline\hline
\T\B 1 & $\overline{u}_n \gamma_k d_n$ & & & & \\
\cline{1-2}
\T\B 2 & $\overline{u}_n \gamma_k d_w$ & $1\;1^{--}$ & $(1,0)\oplus (0,1)$ &
$1\;1^{--}$ & \\
\cline{1-2}
\T\B 3 & $\overline{u}_w \gamma_k d_w$ & & & & \\
\hline
\T\B 4 & $\overline{u}_n \gamma_k \gamma_4 d_n$ & & & & \\
\cline{1-2}
\T\B 5 & $\overline{u}_n \gamma_k \gamma_4 d_w$ & $1\;1^{--}$ &
$\left(\frac{1}{2},\frac{1}{2}\right)_b$ & $1\;1^{--}$ & \\
\cline{1-2}
\T\B 6 & $\overline{u}_w \gamma_k \gamma_4 d_w$ & & & & \\
\hline
\T\B 7 & $\overline{u}_{\partial_k} d_n$ & & & & \\
\cline{1-2}
\T\B 8 & $\overline{u}_{\partial_k} d_w$ & \rb{$1\;1^{--}$} &
\rb{$\left(\frac{1}{2},\frac{1}{2}\right)_b$} & \rb{$1\;1^{--}$} & \\
\hline
\T\B 11 & $\overline{u}_{\partial_i} \gamma_k d_{\partial_i}$ & $1\;1^{--}$ &
$(1,0)\oplus (0,1)$ & $1\;1^{--}$ & \\
\hline\hline
\T\B 1 & $\overline{u}_n \gamma_k \gamma_5 d_n$ & & & & \\
\cline{1-2}
\T\B 2 & $\overline{u}_n \gamma_k \gamma_5 d_w$ & $1\;1^{++}$ & $(1,0)\oplus (0,1)$
& $1\;1^{++}$ & \\
\cline{1-2}
\T\B 3 & $\overline{u}_w \gamma_k \gamma_5 d_w$ & & & & \\
\hline
\T\B 11 & $\overline{u}_{\partial_i} \gamma_k \gamma_5 d_{\partial_i}$ &
$1\;1^{++}$ & $(1,0)\oplus (0,1)$ & $1\;1^{++}$ & \\
\end{tabular}
\end{ruledtabular}
\end{center}
\caption{
List of our meson interpolators. The numbers in the first column together with the quantum numbers $IJ^{PC}$
given in the third column label the interpolators uniquely. The fourth and
fifth column specify the chiral representation and the coupling in the chiral limit. The lattice interpolators may also couple to states
with higher angular momentum $J$ \cite{Lacock:1996vy,Dudek:2007wv}. The subscripts $n$ and $w$ refer to narrow and 
wide smearing of $u$ and $d$ quarks. 
The subscript $\partial_i$ denotes derivative smearing in
$i$-direction. Where it appears, the index 
$i$ is summed over the spatial directions 1,2,3.
The time direction is 4 and the corresponding Dirac matrix
is $\gamma_4$. For the vector and pseudovector channels the  
index $k$ (not summed !) can have values $k = 1,2,3$.
\label{interpolatortable}}
\end{table*}

\subsection{Meson interpolators}

Table \ref{interpolatortable} shows our interpolators for the different meson
channels considered. In the first column the interpolators are numbered according to their
structure rather than consecutively. Numbers 1-6 denote the Jacobi-smeared
interpolators of \cite{Burch:2006dg}, while the interpolators 
7-12 contain at least one derivative. Notice also that only the combination of
number and quantum numbers uniquely labels an interpolator. 

In some cases, an (anti-) symmetrization of the 
interpolators is necessary to obtain the
correct behavior under charge conjugation. Therefore, interpolators denoted as
$\bar{u}_{\partial_i}\Gamma d_{n/w}$ in Table 
\ref{interpolatortable} should be read as 
$\bar{u}_{\partial_i}\Gamma d_{n/w}-\bar{u}_{n/w}\Gamma d_{\partial_i}$.
We restrict ourselves to light, isovector ($I=1$) mesons with degenerate
quark masses $m_u=m_d$.

All interpolators have been classified by their continuum quantum numbers $I\;J^{PC}$,
both for non-vanishing quark mass and in the chiral limit. As usual, $P$ is
the spatial parity, $J$ is the total spin, and $I$ the
isospin. For a neutral $q\bar{q}$ system \footnote{Notice that interpolators
  with covariant derivatives may receive additional terms in their chiral representation.} the $C$--parity is related to the other quantum
numbers in a standard way. To simplify the notation, we omit the spin and
isospin projections in the notation.

In addition to the quantum numbers listed in table \ref{interpolatortable}, the lattice
interpolators will also couple to continuum states with higher $J$ due to the
loss of rotational symmetry \cite{Lacock:1996vy,Dudek:2007wv}. The lattice interpolators 
for $0^{PC}$ mesons couple also to $J\geq 4$, but this does not 
influence our conclusions since there are no observed resonances with 
$J\geq 4$ in the energy regime of interest.  The lattice interpolators 
for $1^{PC}$ mesons couple also to $J\geq 3$ and the issue is 
discussed in section \ref{vectors}.

In the chiral limit, the different interpolators listed in Table
\ref{interpolatortable} can be classified into representations of chiral
$SU(2)_L \times SU(2)_R$ and $U(1)_A$ groups \cite{Glozman:2003bt,Glozman:2007ek}, as well as
with respect to their partial wave $^{2S+1}L_J$ decomposition \cite{Glozman:2007at}.
Below we review these properties. 

We label with $R$ the index of the chiral representation; $R=$ $(0,0)$,
$(1/2,1/2)_a$, $(1/2,1/2)_b$, or $(0,1)\oplus(1,0)$.
The chiral basis $\{R;IJ^{PC}\}$ is obviously consistent with Poincar{\'e} invariance.  
Each of the interpolators in Table \ref{interpolatortable} has a fixed $U(1)_A$ transformation property. Namely, all those interpolators
that belong to $(0,0)$ or $(0,1)\oplus(1,0)$ of $SU(2)_L \times SU(2)_R$ are scalars with respect to $U(1)_A$,
i.e., they transform into themselves under a $U(1)_A$ transformation. However, interpolators with 
opposite spatial parity and the same spin $J$ and isospin $I$ from the distinct $(1/2,1/2)_a$ and $(1/2,1/2)_b$ representations of
$SU(2)_L \times SU(2)_R$ transform into each other upon $U(1)_A$.

The set of quantum numbers  $\{R;IJ^{PC}\}$ uniquely fixes a partial wave content $| I;^{2S+1}L_J\rangle$ of the
quark-antiquark system in the center-of mass frame \cite{Glozman:2007at}. In particular, the different 
interpolators from Table~\ref{interpolatortable} with quantum numbers $1
~0^{++}$ belonging to $(1/2,1/2)_b$ represent the $|1; ^3P_0\rangle$ partial wave in the $\bar q q$ system, 
irrespective of the number of derivatives in the interpolator. Note however
that there are some interpolators in the $1\;0^{++}$ channel (number seven and
eight) which will not couple to the scalars in the chiral limit, since they
belong to the $(0,1)\oplus(1,0)$ representation which requires $J\ge 1$.
 
 For the $1 ~ 0^{-+}$ sector, there are two types of interpolators:
 interpolators (1-3, 9-10, 11) which transform as $(1/2,1/2)_a$ and
 represent the $|1; ^1S_0\rangle$ partial wave and time components of
 pseudovector interpolators which couple to pseudoscalars due to PCAC and which belong to the $(0,1)\oplus(1,0)$
 representation. 

In the $1\;1^{++}$ sector, all  interpolators transform as $(0,1)\oplus(1,0)$ and
 couple only to the $| 1; ^3P_1\rangle$ partial wave. 
 
 However, there are two kinds of interpolators with quantum numbers $1 ~ 1^{--}$. 
 They are the {\it fixed} and orthogonal superpositions of two different partial waves:

\begin{eqnarray}
\displaystyle |(0,1)+(1,0);1 ~ 1^{--}\rangle&=&\sqrt{\frac23}|1;{}^3S_1\rangle+\sqrt{\frac13}|1;{}^3D_1\rangle,\nonumber\\
\displaystyle |(1/2,1/2)_b;1 ~ 1^{--}\rangle&=&\sqrt{\frac13}|1;{}^3S_1\rangle-\sqrt{\frac23}|1;{}^3D_1\rangle.\nonumber
\end{eqnarray} 

The interpolators 1-3 and 11 from Table \ref{interpolatortable} belong to the $|(0,1)+(1,0);1 ~ 1^{--}\rangle$ representation,
while all others transform as $|(1/2,1/2)_b;1 ~ 1^{--}\rangle$.

\subsection{Technicalities}

For our analysis we used 99 uncorrelated quenched gauge
configurations generated with the L\"uscher-Weisz
gauge action \cite{Luscher:1984xn,Curci:1983an}. 
We work on a $16^3\times 32$ lattice with
$a=0.148\ \mathrm{fm}$ determined \cite{Gattringer:2001jf} from the Sommer
parameter (using $r_0 = 0.5$ fm). 
For comparison we also use old data from a $20^3 \times 32$ lattice at 
$a = 0.119$ fm, where, however, only the Jacobi-smeared sources without
additional derivatives are available. The boundary conditions for the gauge
fields are periodic in all four directions.
The quark propagators were computed from the Chirally Improved 
Dirac operator \cite{Gattringer:2000js,Gattringer:2000qu} with periodic
boundary conditions in space and antiperiodic boundary conditions in the time coordinate. We study
several quark mass parameters in the range $am_q=0.02\dots 0.2$. We fold
individual entries of the correlation matrix resulting from propagation in
positive and negative time direction according to their symmetry, which
reduces the statistical errors and improves the quality of the data
significantly. Unless noted
otherwise, the errors we quote are statistical errors determined with the
Jackknife method. 

Where possible we also indicate the systematical uncertainties by a shaded
band (Figs.~\ref{fig_pion_first}, \ref{fig_pion_second} etc.). The upper and
lower limits of this band are obtained by repeating the fits of the
eigenvalues using different fit ranges and varying the interpolators used in
the correlation matrix. Although this is certainly only a rough estimate of
the systematic uncertainty in the mass determination, we refer to these error
estimates as ``systematic errors''. We stress, however, that these
errors do not include the error introduced by the quenched
approximation. 

For plots with the pion mass squared on the horizontal axis, 
we use a specific combination of gaussian interpolators for extracting the ground
state pion mass. The corresponding statistical error gives rise to the 
horizontal error bars in some of our plots which are, however, 
smaller than the symbols used.

When fitting eigenvalues obtained with the variational method we experimented
with both correlated and uncorrelated (two parameter) exponential fits. For
the correlated fits we used a jackknife estimate of the correlation matrix
which was not always stable with our ensemble of configurations. For the
generation of the plots we therefore resorted to simple uncorrelated fits throughout,
using only the diagonal elements of the covariance matrix.
For the cases, where correlated fits are stable and a direct comparison with
uncorrelated fits is possible we find that the latter give larger statistical errors
\cite{Gattringer:2003qx}. This explains the rather small
$\chi^2/\mathrm{d.o.f.}$ we find. Thus the uncorrelated errors we give in the plots and in the
appendix are probably overestimated. 

\subsection{A first look at the derivative sources}

Figure \ref{fig_distribution} shows the diagonal elements of the correlation
matrix for some of the channels considered. Compared to the interpolators from
reference \cite{Burch:2006dg} (without derivative sources), 
the interpolators with derivative sources show
stronger contributions from excited states, i.e., they have a steeper slope
for small Euclidean time $t$. Nevertheless, for all interpolators
the ground state in the respective channel dominates the behavior at large 
time separation, i.e., all correlators in Fig.~\ref{fig_distribution} display
the same slope at sufficiently large $t$. 

We remark that the propagators in the $0^{++}$ channel at the lightest quark masses show a
deviation from that pattern for small $t$ due to ghosts. A more
detailed discussion of the scalar correlators will be given in Section
\ref{scalars}.

As suggested by Fig.~\ref{fig_distribution}, the pion ground state can be fitted
from a single diagonal correlator at all quark masses. Stable plateaus are
obtained in the time interval $t=6\dots 15$ where cosh-fits can be
performed for the individual correlators. 
For the lowest quark mass $am_q=0.02$ the statistical error usually is of the
order of $1-2\%$ of the fitted value with the derivative 
sources leading to a somewhat larger
error. An exception is interpolator 12 where the statistical error is about
$3.5\%$ of the fitted value. At larger masses the error is substantially
smaller. The fit values for the ground state pion mass from the different
individual correlators, shown in the lhs.~plot of Fig.~\ref{fig_distribution},
agree within two sigma.

For the $1^{--}$ channel similar observations hold. 

\begin{figure}[tb]
\includegraphics[width=8cm,clip]{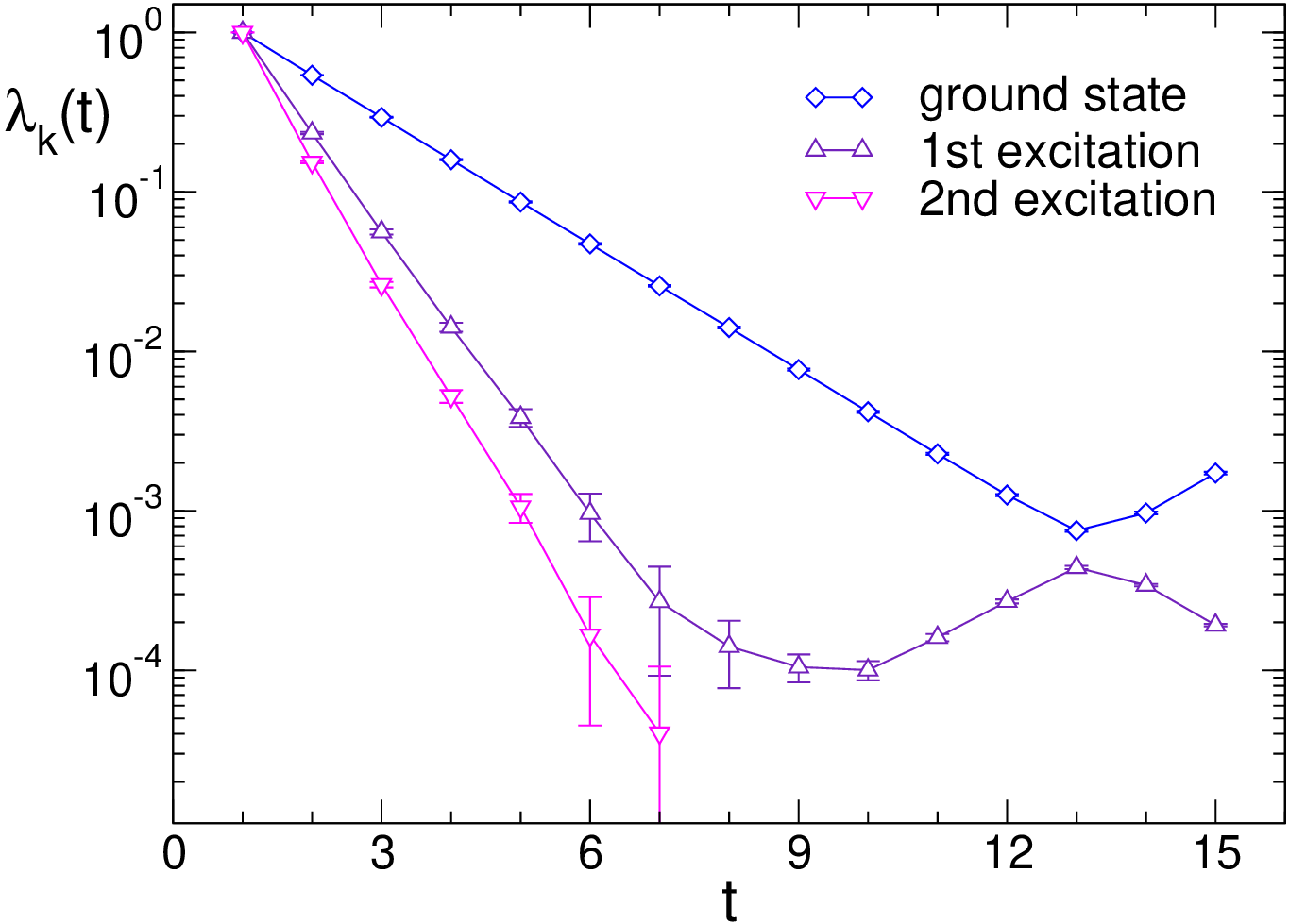}
\caption{First three eigenvalues from the generalized eigenproblem 
(\ref{generalized}) for the $0^{-+}$ channel as a function of 
$t$.}
\label{fig_evals}
\end{figure}

\subsection{Contributions from backward propagation}

In Fig.~\ref{fig_evals} we show the three largest eigenvalues obtained 
from the generalized problem (\ref{generalized}) for the pion case. 
It is obvious that on the
logarithmic scale used in this plot the three eigenvalues give rise to three 
essentially straight lines for sufficiently small $t$. The different slopes 
correspond to the masses of the ground-, first- and second excited states. 

For mesons, forward and backward propagation behaves in the same way. Each
interpolator coupling to a particular state at early times will also couple to
the same, but backward running, state at later time. Higher channels in the
generalized eigenvalue problem may in this way turn into lighter state signals.
This interesting observation (see also \cite{Lichtl:2007jc}), particular for the generalized eigenvalue problem,  
can be made for the second eigenvalue in Fig.~\ref{fig_evals}: At
$t\sim 9$ (when using $t_0=1$)
the data points of the second eigenvalue
change their behavior and start to increase again. 
Beyond $t=9$ the data points form a straight line with positive slope. This
upward pointing straight line turns around again at $t \sim 13$ and from
there on decays with a slope corresponding to the ground state mass. Also the
slope of the upwards pointing piece between $t = 9$ and $t = 13$ (which is
then continued by an upward pointing piece of the largest eigenvalue), has the
slope of the ground state mass. 

This avoided level crossing scenario has an important consequence: The
generalized eigenvalue problem disentangles the forward propagating ground and excited state
masses only up to the first crossing with the backward running lightest propagator,
which for our example happens at $t \sim 9$. Beyond that value also the
second eigenvalue, which for small $t$ is dominated by the first excitation,
couples to the lighter ground state (running ``backwards'') 
and no longer provides information on  
the excitations. In particular in the $0^{-+}$ channel, where a backward
running light pion crosses with the second eigenvalue already at a small $t$, this
effect limits the analysis of excited states \footnote{We remark, however, that
  in a 2-d pilot study, where very accurate data can be obtained, we found
  that after the first crossing the third eigenvalue takes over the slope of
  the first excitation. Thus it seems possible to extract additional
  information on the first excitation from higher eigenvalues.}.

The comments of this subsection are particularly important for light
pseudoscalars. For the pions it is the backward running contributions that
limit the fit range with the generalized eigenproblem, leading to errors
comparable with the errors obtained from fitting single correlators with the
correct functional form.

\begin{figure}[tb]
\includegraphics[width=7.8cm,clip]{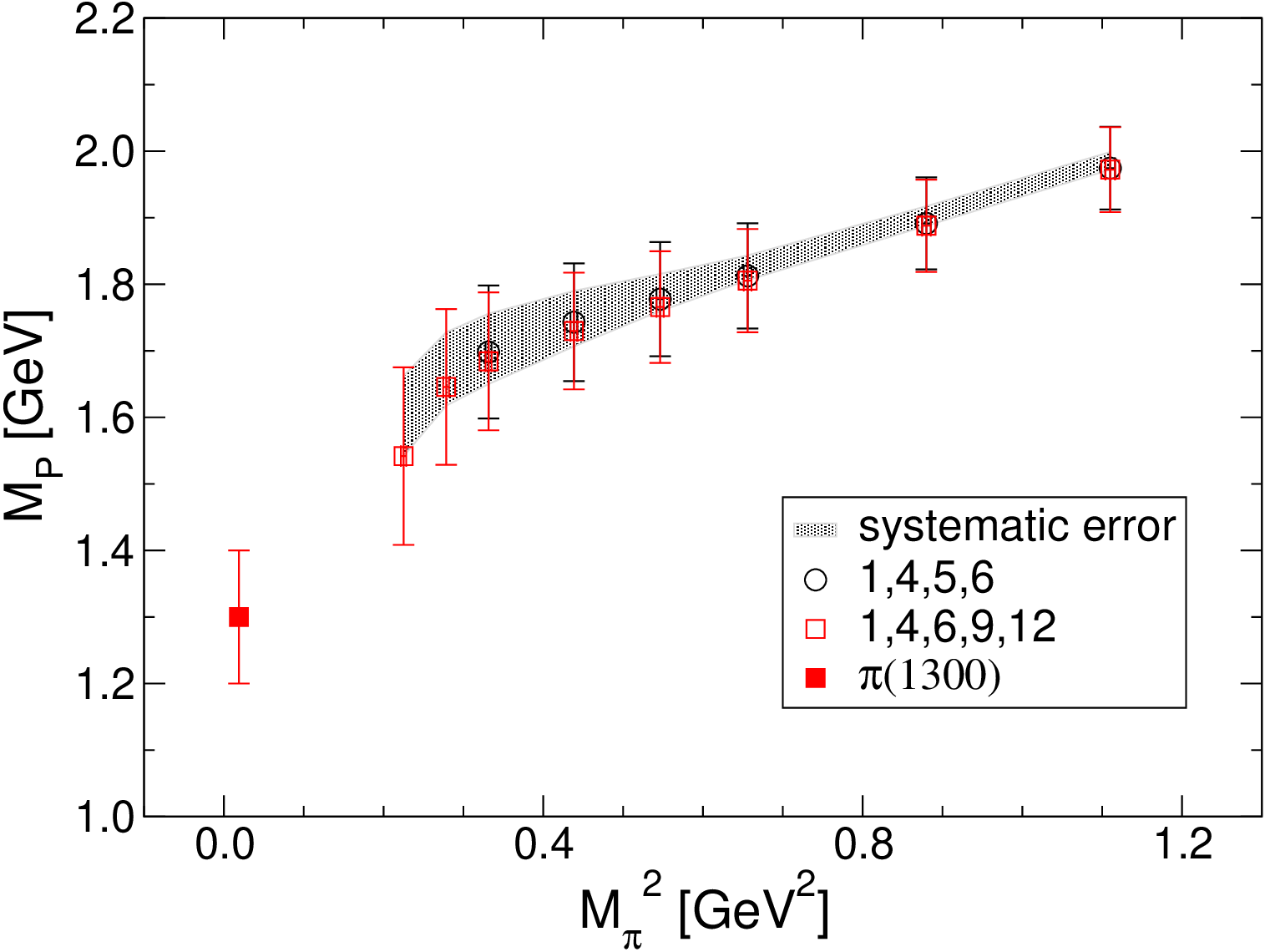}

\caption{First excited state of the pseudoscalars for two different sets of 
interpolators. The
error bars are statistical only and the shaded region indicates the additional
systematic error as discussed. 
The filled symbol corresponds to the experimentally measured $\pi(1300)$.
}
\label{fig_pion_first}
\end{figure}

\begin{figure}[tb]
\includegraphics[width=7.8cm,clip]{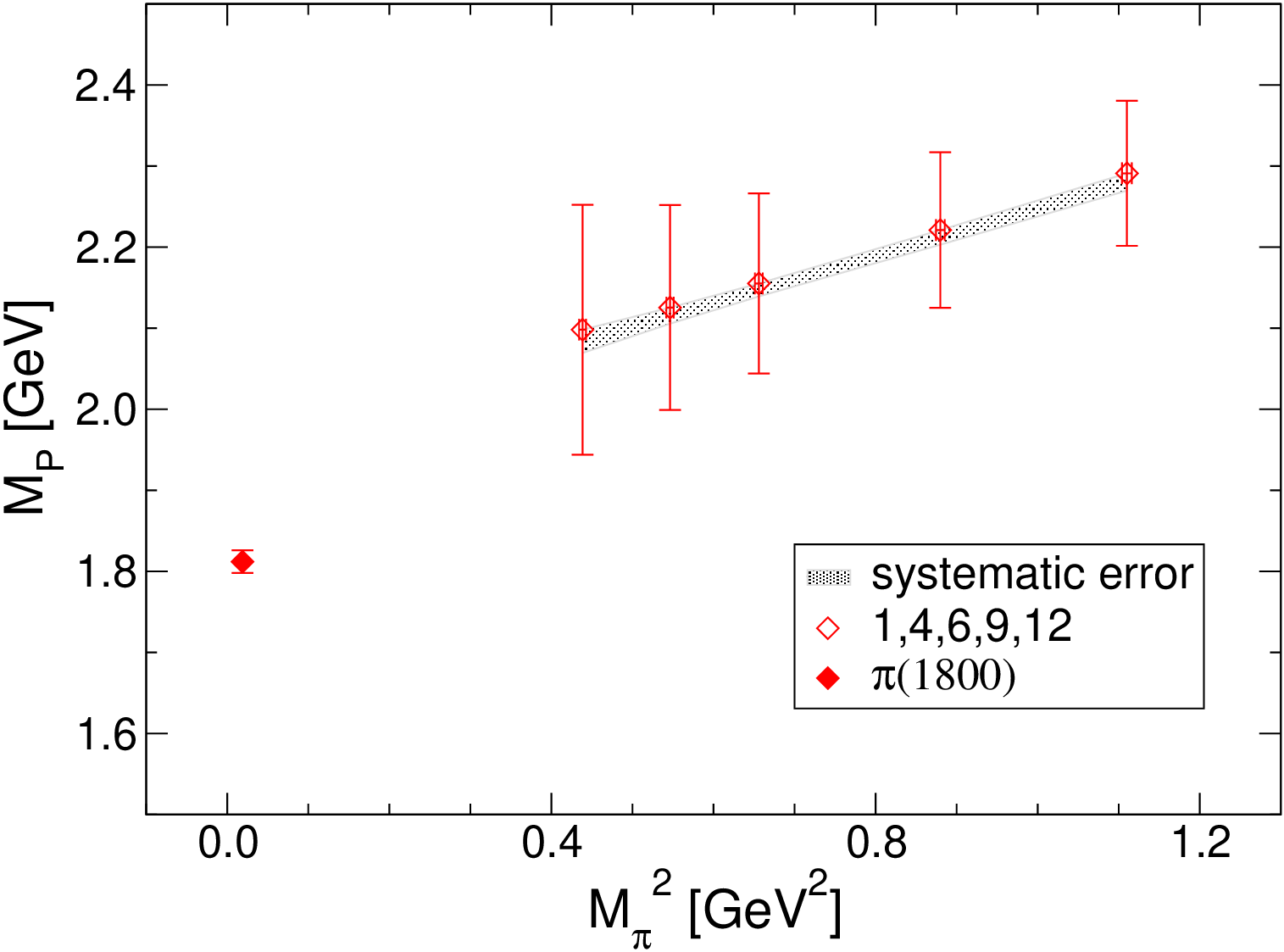}

\caption{Second excited state for the pseudoscalars. The
  filled symbol indicates the $\pi(1800)$. 
}
\label{fig_pion_second}
\end{figure}

\begin{figure*}[bt]
\begin{center}
\includegraphics[scale=0.28,clip]{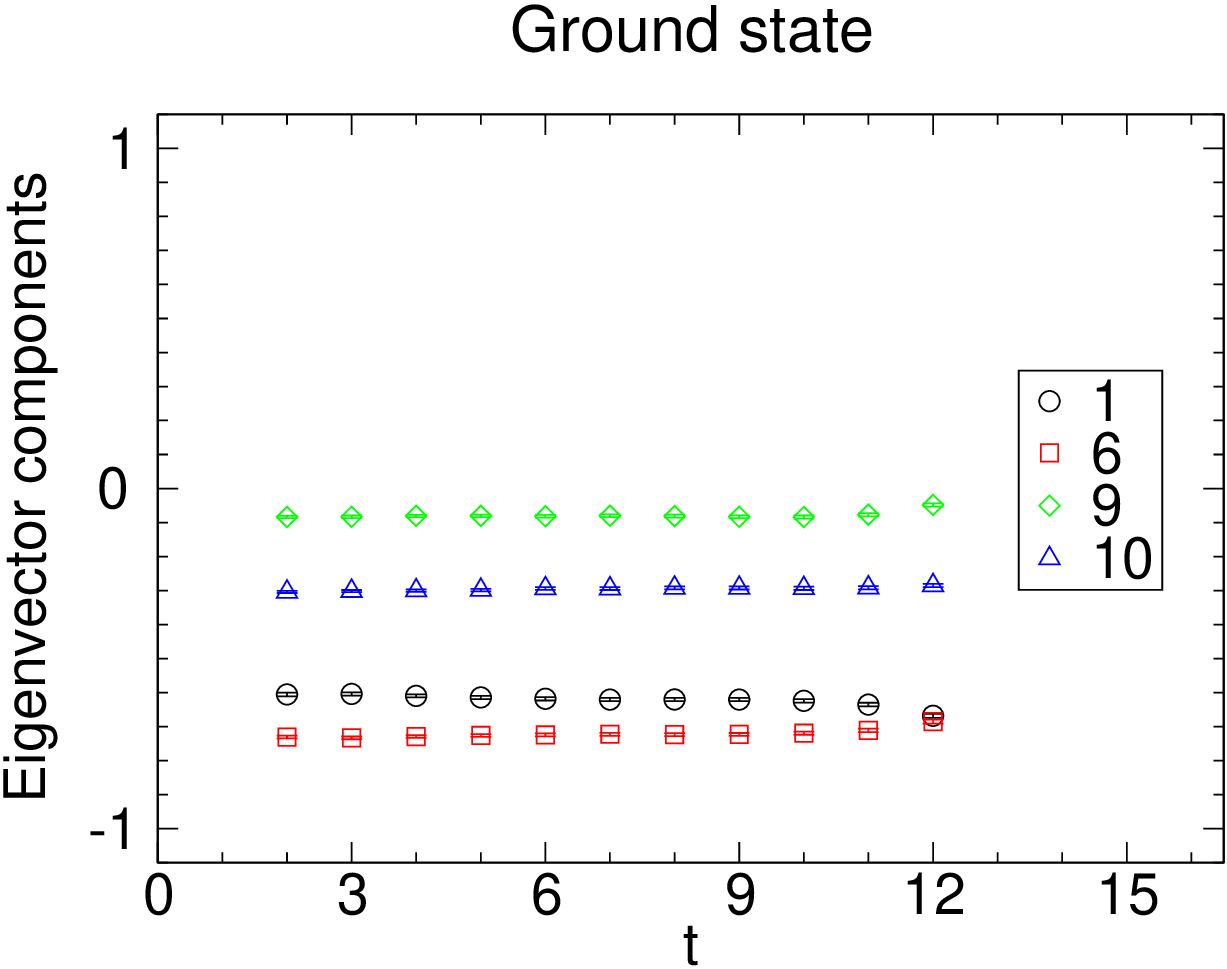}
\hspace{0.2cm}\includegraphics[scale=0.28,clip]{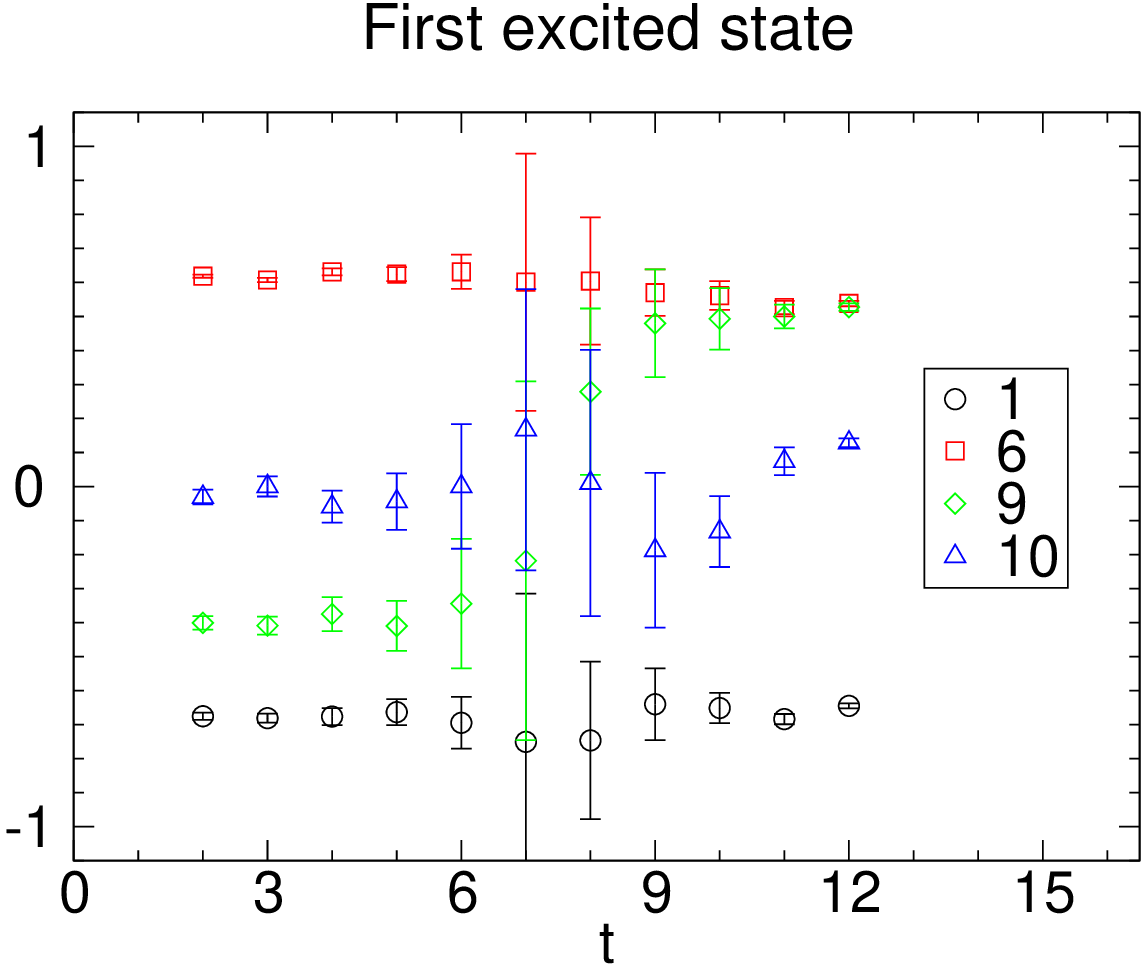}
\hspace{0.2cm}\includegraphics[scale=0.28,clip]{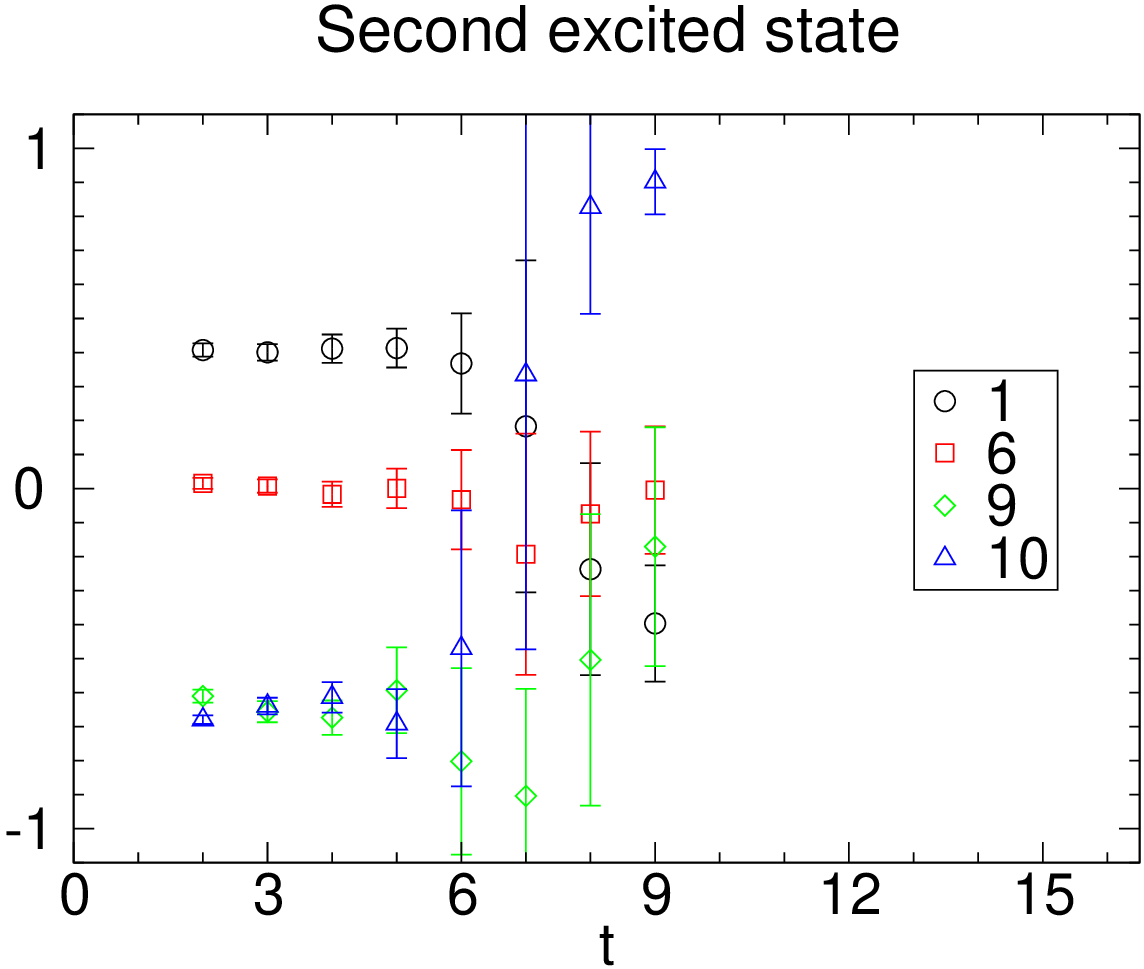}
\end{center}
\vspace*{-2mm}
\caption{Eigenvector components from the standard eigenvalue problem 
(\ref{normalev}) as a function of $t$. The eigenvectors correspond to
the pion ground state (lhs.~plot), the first and second excited states (center and
rhs.~plots) and are obtained from a $4\times 4$ matrix of the
interpolators 1, 6, 9, 10, at quark mass $am_q = 0.12$.}
\label{fig_pion_eigenvectors}
\end{figure*}

\section{Results for individual meson channels}

\subsection{The $0^{-+}$ channel}

Figure \ref{fig_pion_first} shows the results for the first excited state of
the pion. We display the results for two different sets of interpolators.
Circles are used for the combination of gaussian interpolators 1, 4, 5, 6, while the squares
correspond to the combination 1, 4, 6, 9, 12 of gaussian and derivative interpolators.
While combinations of gaussian and derivative interpolators allow for fits at
slightly lighter quark masses, the interpolators with derivative sources
couple weaker to the ground state. The systematical
uncertainty (shaded region in the figure) from the choice of interpolators is
consistent with statistical effects as reasonable combinations of four or more
interpolators lie within one sigma of our final fit result. 
While there is no significant improvement,
the new results with a larger basis nicely confirm the existence of the
measured state.

With a combination of gaussian and derivative interpolators, it is also possible to obtain
fits for a second excited state which could not be observed before. This state
is displayed in Fig. \ref{fig_pion_second}. In the
chiral limit, this state can most likely be identified with the
$\pi(1800)$. Fits with various different combinations of interpolators lead
to the same results which all show stable eigenvector entries.

It is instructive to look at the components of the eigenvectors for all three
states observed in the pseudoscalar channel. Figure
\ref{fig_pion_eigenvectors} shows such a plot. While the derivative
interpolators 9 and 10 do not contribute significantly to the ground and
first excited states, they are most important for obtaining the newly observed
second excited state. This behavior is qualitatively the same for all possible
combinations of interpolators where the second excited state could be seen.

While the reduction in the statistical error for the first excited
state can merely be attributed to an enlargement of the basis, the second
excited state is only observed when including derivative interpolators. We
would like to stress that a correlation matrix of similar size consisting
solely of non-derivative operators does not enable us to see this excitation.

\subsection{\label{scalars}The $0^{++}$ channel}

\begin{figure}[b]
\includegraphics[width=8cm,clip]{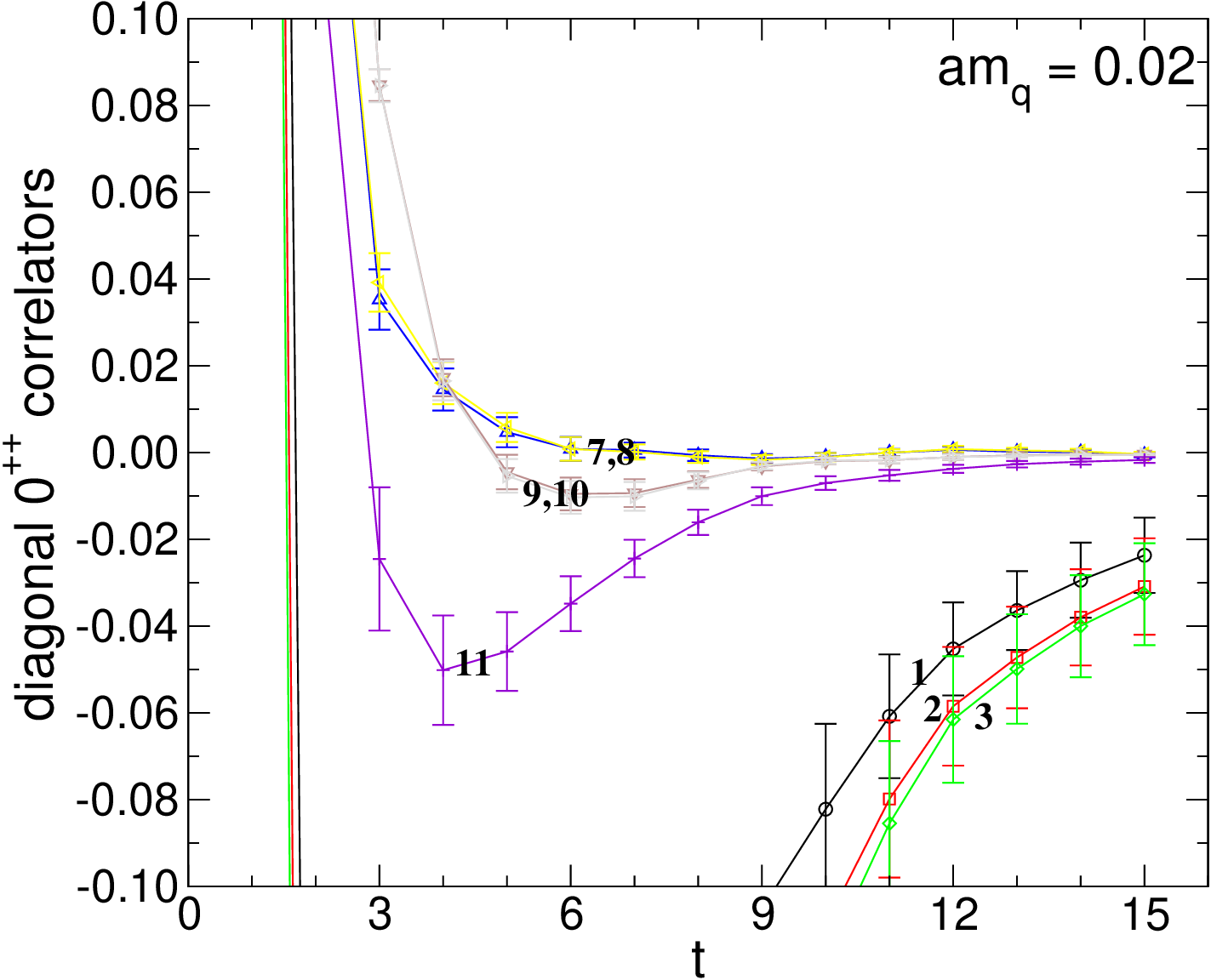}
\caption{
Diagonal entries of the correlation matrix for the scalar channel at $am_q =
0.02$. The numbers next to the data label the correlators according to Table I. 
There are clear contributions of ghost states for
some interpolators (i.e., negative values),
while other interpolators seem to be almost free of them. 
}
\label{fig_ghostcontribution}
\end{figure}

In the $1\;0^{++}$ channel contributions from ghosts
\cite{Bardeen:2001jm,Bardeen:2003qz,Burch:2005wd,Prelovsek:2002qs,Prelovsek:2004jp,Mathur:2006bs}, arising from the $\eta\prime \, \pi$ 
contribution to the isovector-scalar correlators, are expected and must
be identified for a clean interpretation of the data. 
These unphysical contributions due to quenching have a negative spectral weight and
dominate the correlators at small quark masses, leading to correlators which
become negative at intermediate time separations.
Figure \ref{fig_ghostcontribution} shows the diagonal correlators for the scalar
channel at our smallest quark mass $am_q=0.02$. While some of them (1-3) display a very
prominent ghost contribution, others show a much smaller contribution.
Correlators 7 and 8 feel no effects from ghosts at all with our limited
statistics, which may be related to their different chiral structure (see
Table \ref{interpolatortable}). We note that these two interpolators do not
seem to couple in the dynamical case, as our preliminary dynamical results do
not show a signal for these correlators.

While the variational approach enables us to disentangle most of these ghost
contributions \cite{Burch:2005wd}, at low quark masses the quality of the
observed plateaus quickly deteriorates for the gaussian interpolators 
leading to large error bars for the $a_0$ ground state.

\begin{figure}[t]
\includegraphics[width=8cm,clip]{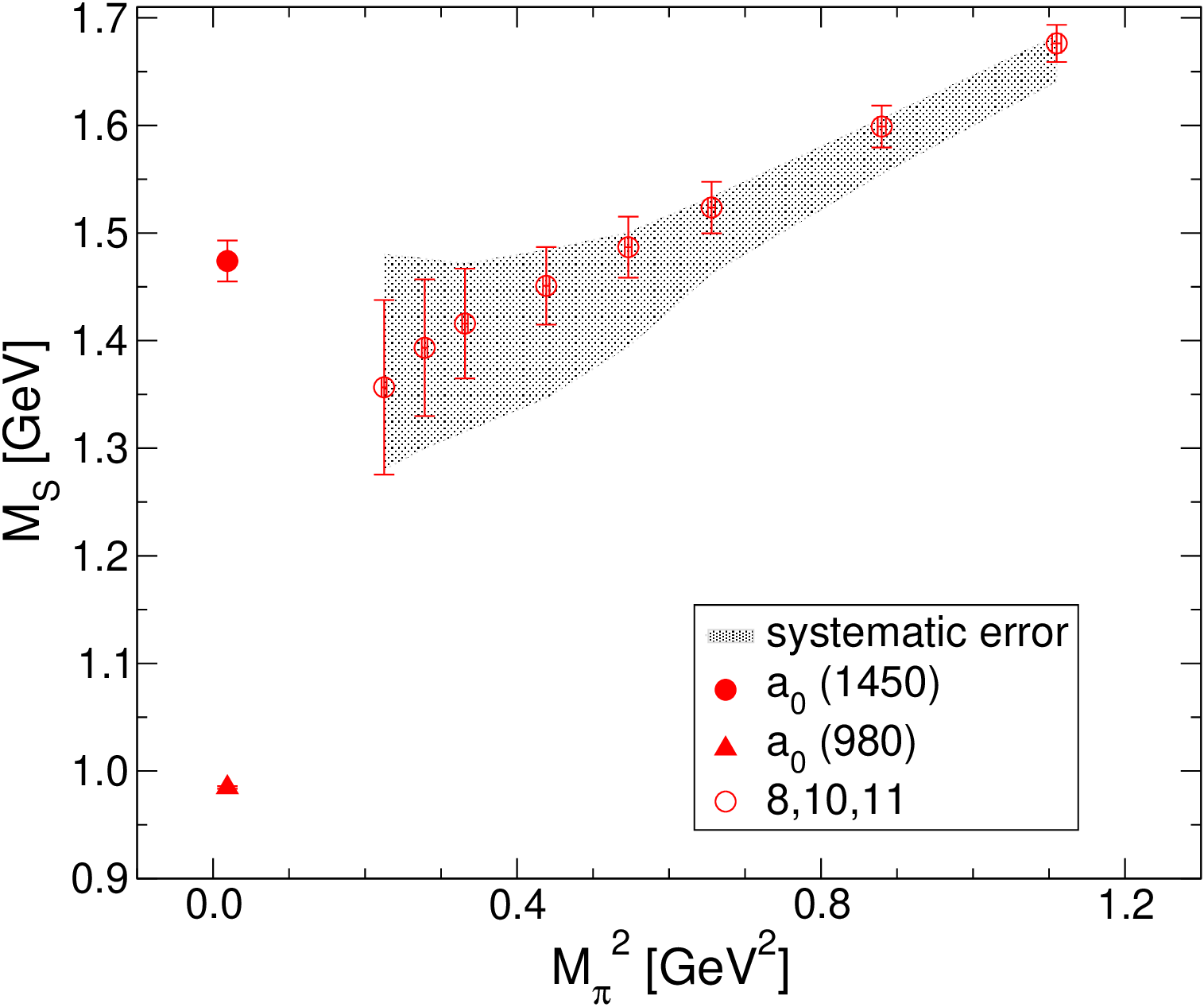}
\caption{
Ground state mass of the isovector, scalar ($a_0$). The
error bars are statistical only and the shaded region indicates the additional
systematic error described in the text. The black circle shows the
(experimentally) measured $a_0(1450)$ and the triangle indicates the $a_0(980)$.
}
\label{scalar_masses}
\end{figure}

The additional interpolators with derivative sources enlarge the correlation
matrix, which is vital in the presence of ghost contributions. Furthermore, it
is these interpolators which couple only weakly to the ghosts. Using the
variational method we are able to disentangle the leading ghost contribution
and in some cases even a sub-leading ghost contribution, corresponding to a $\eta^\prime \, \pi$ state with relative momentum.

Figure \ref{scalar_masses} shows the results for the largest eigenvalue of the
variational analysis. The plot demonstrates that derivative sources enable us to perform fits at smaller quark masses with reduced
statistical errors in the intermediate and heavy quark mass region. 

The shaded
region in Fig.~\ref{scalar_masses} indicates our estimate of the additional 
systematic errors
due to the choice of fit ranges and interpolators considered. As it remains
unclear which systematics cause the dependence on the choice of interpolators
in this channel,
we refrain from an extrapolation to the physical mass region. Interpolators
with strong ghost contributions, however, tend to lead to higher mass values
suggesting the ground state to be in the region of the $a_0(1450)$, while a
fit with those interpolators containing no visible ghost contribution leads to
values in the lower parts of the shaded region. 

For this channel large
quenching effects might influence the result. Thus it is not clear whether the
data should extrapolate to the $a_0(980)$ or the $a_0(1450)$. 
More light will be shed on this
channel only when dynamical data are available where no quenched ghosts 
are present \cite{Frigori:2007wa}. 
If the systematic deviations we observe in the choice of 
interpolators are due to
ghosts, one should expect the results to become more consistent with dynamical
data.

A review of issues faced in the scalar channel and a detailed
discussion of the possible nature of the isovector scalar groundstate can be
found in the recent review by McNeile \cite{McNeile:2007fu}.

\begin{figure}[t!!!]
\includegraphics[width=8cm,clip]{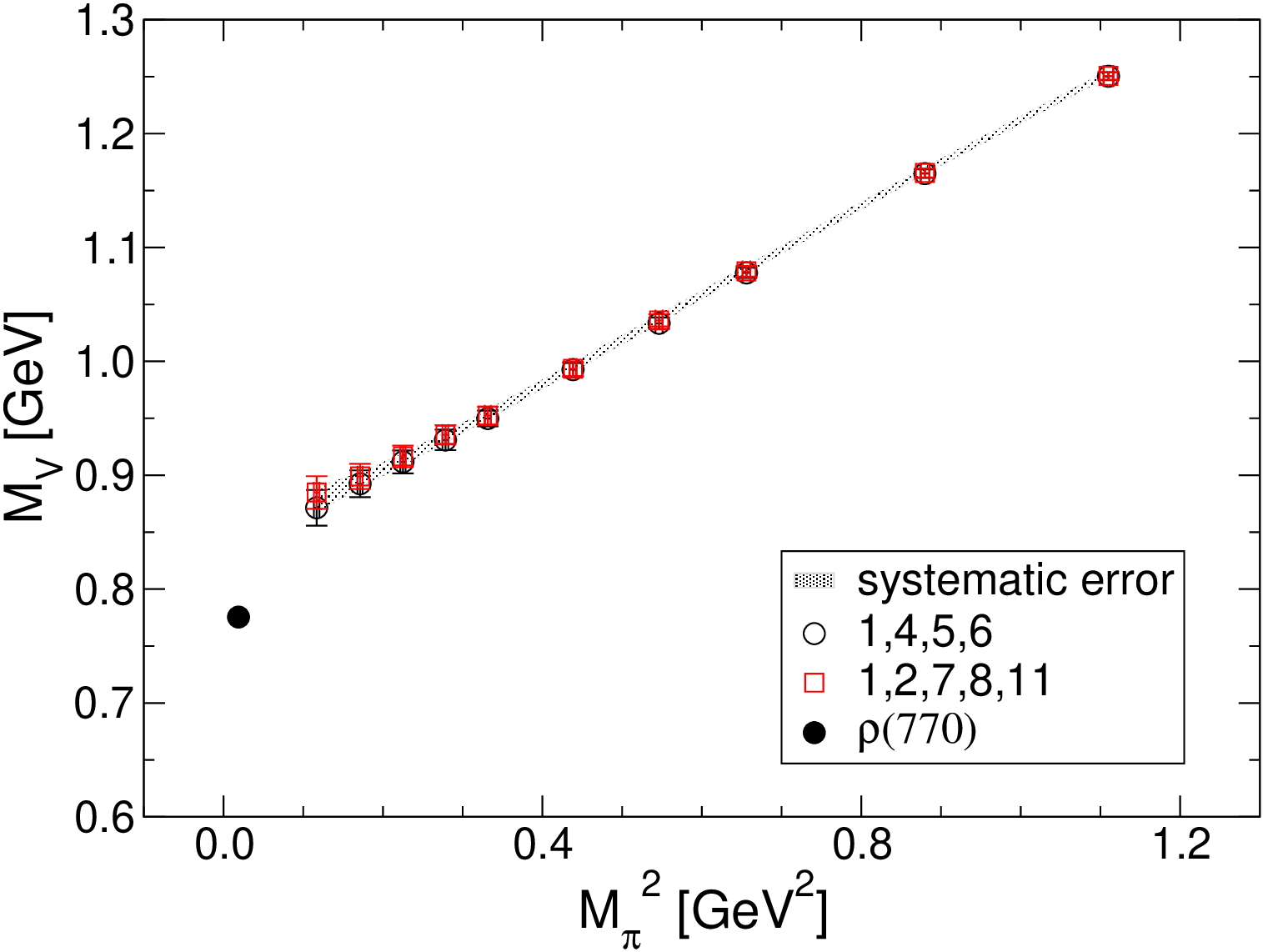}
\vspace{-1mm}
\caption{
Ground state for the $\rho(770)$ meson.
}
\label{fig_vector_groundstate}

\vskip4.5mm

\includegraphics[width=8cm,clip]{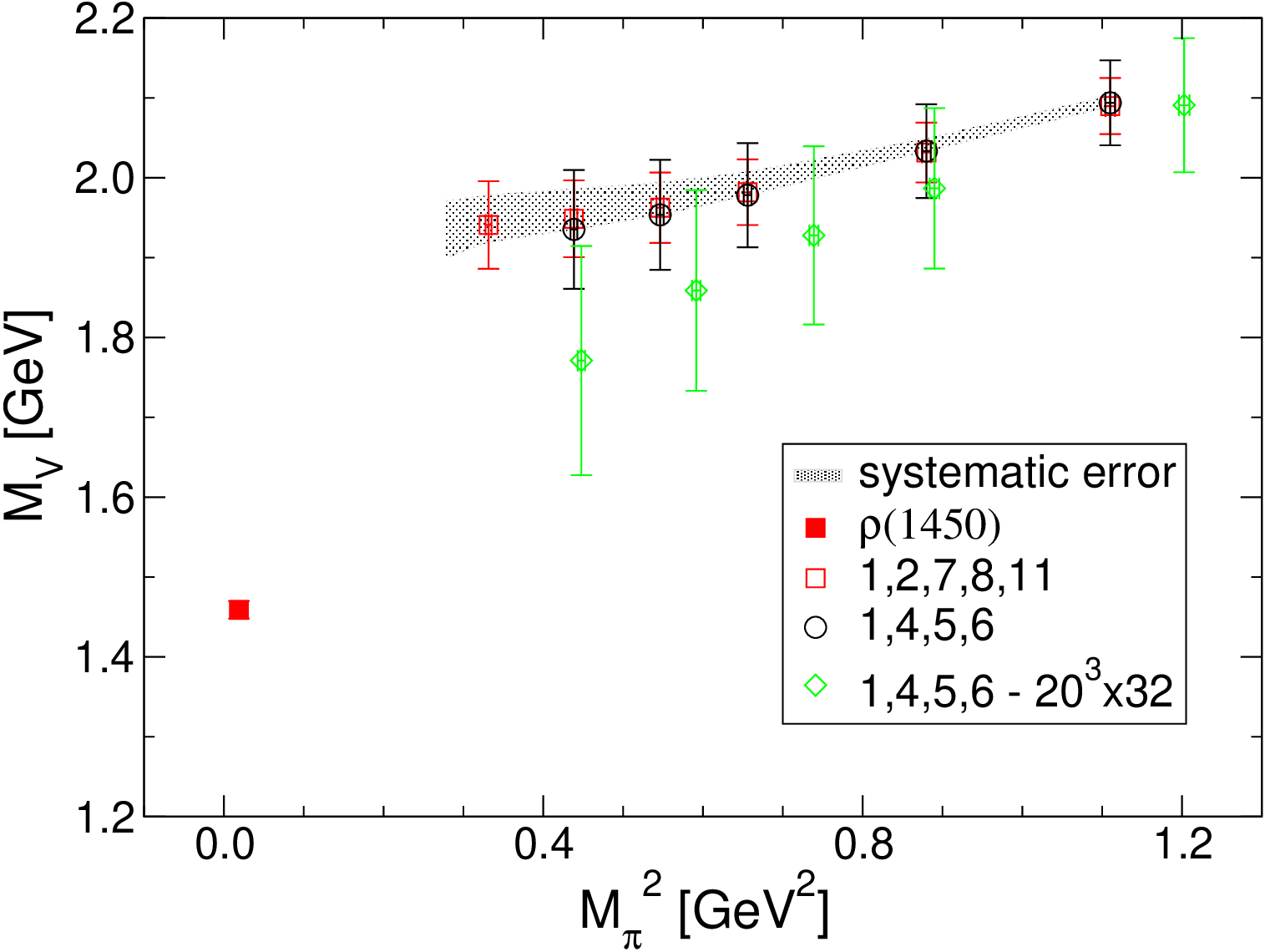}
\vspace{-3mm}
\caption{
First excited state for the $\rho$ meson, compared to the
first experimental meson resonance with $J=1$, the $\rho(1450)$. 
The diamonds represent an alternative fit  
of data from \cite{Burch:2006dg} for 
a finer lattice of the same volume ($20^3\times 32$, $a = 0.119$ fm).
}
\label{fig_vector_first}

\vskip4.5mm

\includegraphics[width=8cm,clip]{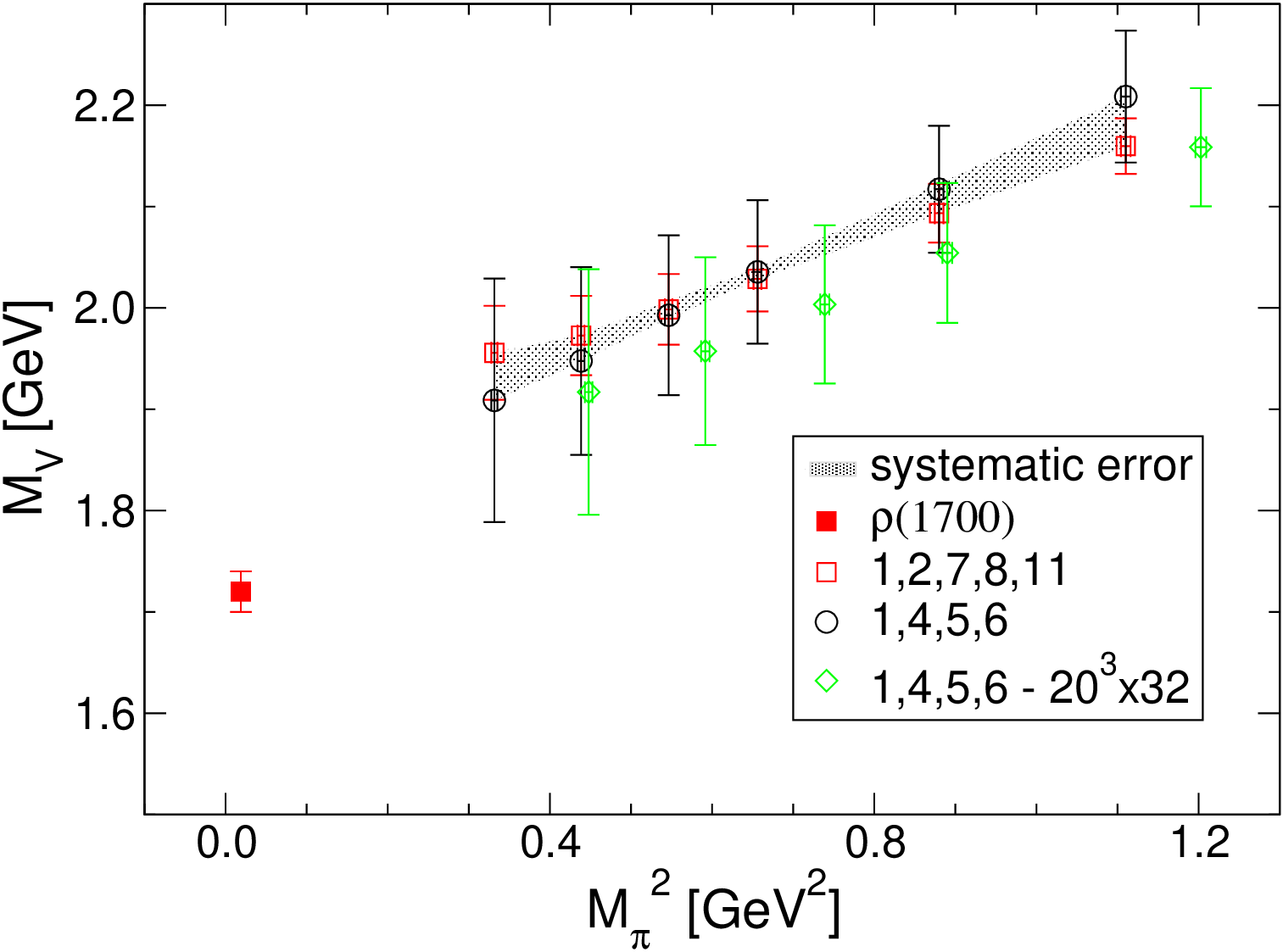}
\vspace{-1mm}
\caption{
Second excitation in the vector channel compared to the $\rho(1700)$. Again
the diamonds indicate data from the finer lattice.
}
\label{fig_vector_second}
\end{figure}

\subsection{\label{vectors}The $1^{--}$ channel}

As mentioned before, the ground state for the vector meson, 
the $\rho(770)$, can also be fit from
single correlators. However the results improve quite drastically
when a matrix of interpolators is used.

Figure \ref{fig_vector_groundstate} shows the $1^{--}$ meson ground state and
illustrates the good quality of the data in the quenched approximation, where
no decay is possible. The results from different interpolators and fit
ranges agree within error bars.

The interpretation for the first and second excitation in the $1^{--}$ channel
is less clear. From experiment we know of multiple excitations with $J=1$ below 2 GeV, the most established
being the $\rho(1450)$ and the $\rho(1700)$. In addition, due to loss of
continuous Lorentz symmetry, some of the lattice
interpolators we chose may in principle couple to continuum states with 
higher $J$ \cite{Lacock:1996vy,Dudek:2007wv},
and there is at least one such excitation known in the vector
channel, the $\rho_3 (1690)$. Excluding this possibility would be a difficult task which might be
overcome by taking a look at different irreducible representations of the
hyper-cubic group where degeneracy of states in different representations
can be used to identify the correct $J$, as has been demonstrated for 
baryons \cite{Basak:2005ir,Basak:2005aq,Basak:2007kj}.

As can be seen in Figs.~\ref{fig_vector_first} and
\ref{fig_vector_second}, the values obtained from combinations of gaussian
interpolators agree qualitatively with the values obtained from the
larger basis, while the larger basis leads to overall smaller error bars and
somewhat more stable plateaus. Nevertheless, the two excitations are too close
together and would both be consistent with the $\rho(1700)$. This problem has
already been encountered in \cite{Burch:2006dg}, where both the $16^3 \times 32$
lattice with spacing $a=0.148$ fm, and a finer $20^3 \times 32$
lattice with $a=0.119$ fm have been used. The data
from the fine lattice lead to two distinct excitations compatible with an
interpretation as the physical $\rho(1450)$ and $\rho(1700)$. To demonstrate
this, we also included an alternative fit of the old data from the fine
lattice in Figs.~\ref{fig_vector_first} and \ref{fig_vector_second}.

Inspecting the eigenvectors of the states, we can identify
them by their operator content and come to the conclusion that they are
indeed the same on both lattices. Moreover, as both lattices have the same physical
volume, this leads us to an interpretation of the difference as a
discretization effect. Such an explanation seems reasonable as the states are
rather close to each other. We therefore are confident that the method works and
that the same calculation on finer lattices would lead to results in better
agreement with experiment.

\subsection{The $1^{++}$ channel}

\begin{figure}[h!]
\vspace{4mm}
\includegraphics[width=8cm,clip]{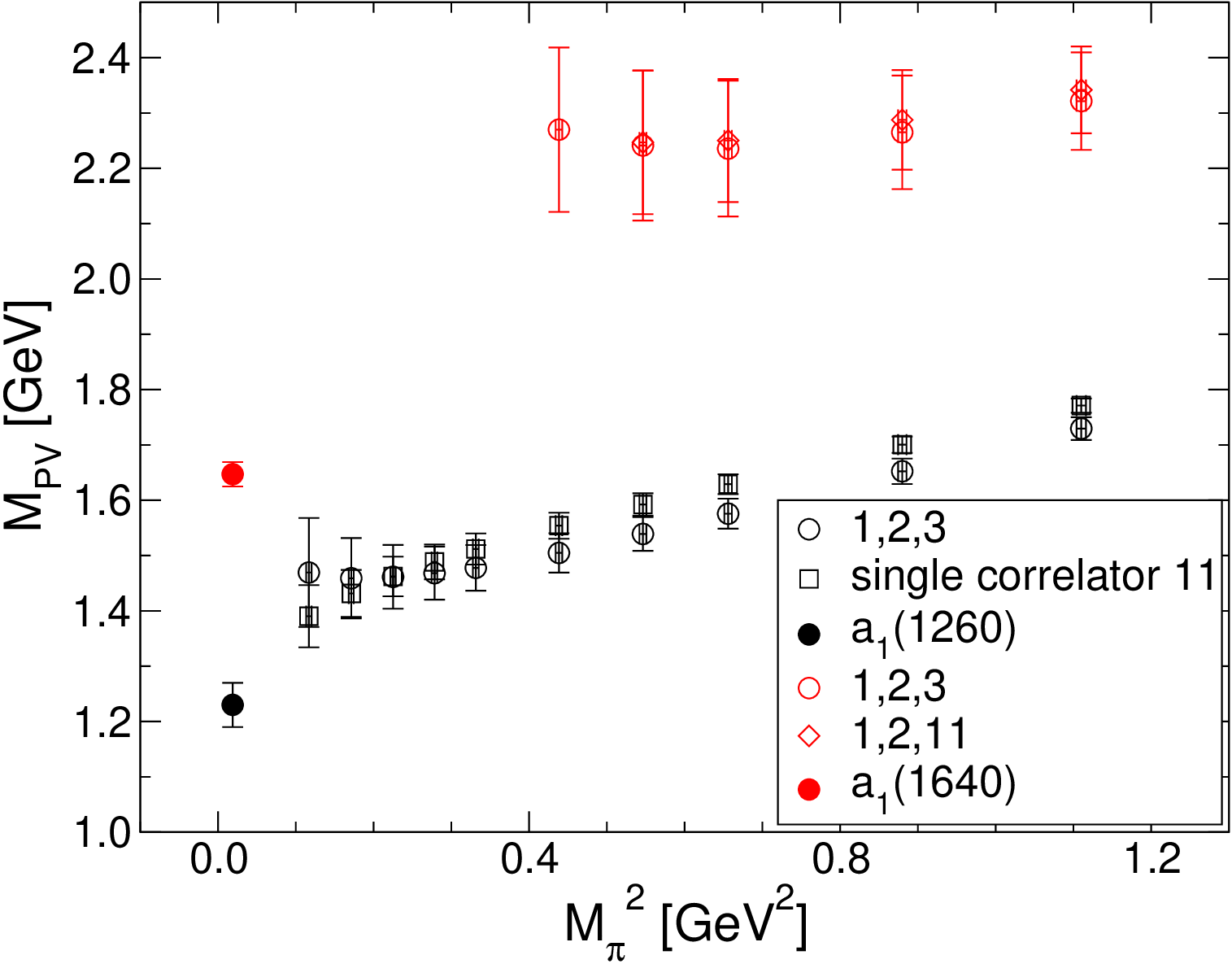}
\caption{
Ground- and first excited state of the pseudovector mesons ($a_1$). 
The filled circles indicate the physical states.
}
\label{fig_axial_extrapol}
\end{figure}

For the $1^{++}$ channel, there is only one interpolator containing derivative
quark sources. Figure \ref{fig_axial_extrapol} shows data obtained from different
combinations of gaussian interpolators and the one containing derivatives. An indication of error bands,
as shown for the other channels, has been omitted here, since 
the two combinations plotted already show the extremes.

Figure \ref{fig_axial_extrapol} demonstrates a clear improvement in the description
of the ground state using the interpolator with derivative quark sources. While
the results from the gaussian and the full sets agree qualitatively, the statistical
errors towards smaller quark masses are significantly reduced. The reason for
this are longer, more stable effective mass plateaus allowing for larger
fit ranges. At larger quark masses there is a slight deviation of the order of
two sigma.

The excited state previously observed stays the same if one includes the new
interpolator in the analysis. Looking at the components of the modified
eigenvalue problem we see that this interpolator contributes only weakly
to this excited state.

\section{Summary}

In this article we have explored the impact of derivative sources in light meson
spectroscopy. The sources are obtained by applying a covariant derivative on
a Jacobi-smeared quark source. Interpolators based on derivative sources were tested in
a quenched excited meson spectroscopy calculation based on the variational
method. 

We find that both, ground- and excited state signals may be improved, depending
on the channel. For the $0^{-+}$ channel we find that the quality of the first
excited state improves and fits become possible for smaller quark
masses. In addition a second excited state can be identified when adding
interpolators with derivative sources in the correlation matrix. 

The $0^{++}$ channel is dominated by the presence of ghosts at small quark
masses which we essentially disentangle using the variational method.
The additional interpolators with derivative sources are helpful since they
enlarge the correlation matrix and some of them couple only very weakly to
the ghost states. They allow one to fit scalar masses at lower quark masses,
although the results still depend significantly on the choice of interpolators
used in the correlation matrix. This indicates that results for this channel
from a quenched calculation should be interpreted only with the necessary
caution.

For the $1^{--}$ mesons we demonstrate that the inclusion of derivative sources
leads to results with a smaller statistical error which agree excellently with
the results published before.

For the $1^{++}$ channel we show that the derivative sources drastically
improve the signal for the ground state. At the same time a matrix of
interpolators enables us to identify a second excited state.

\begin{acknowledgments}
We thank Tommy Burch, Julia Danzer and Markus Limmer for discussions.
D.M.~and L.Ya.G.~acknowledge support by ``Fond zur F\"orderung der
wissenschaftlichen Forschung in \"Osterreich'' (FWF DK W1203-N08 and
P19168-N16). The data were generated on computers at the Leibniz
Rechenzentrum, Garching and at the ZID, Graz.  
\end{acknowledgments}

\appendix*

\section{Tables}

\begin{table}[tbh]
\begin{center}
\begin{ruledtabular}
\begin{tabular}{c|c|c|c|c|c}
interpolators & $m_q$ & mass [GeV] & error & fitrange & $\chi^2/\mathrm{d.o.f.}$ \\
\hline\hline
\T\B 1,4,5,6 & 0.20 & 1.0535 & 0.0025 & 4-10 &  0.09\\
\cline{2-6}
\T\B & 0.16 & 0.9379 & 0.0025 & 4-10 &  0.11\\
\cline{2-6}
\T\B & 0.12 & 0.8098 & 0.0026 & 4-10 &  0.13\\
\cline{2-6}
\T\B & 0.10 & 0.7391 & 0.0027 & 4-10 &  0.14\\
\cline{2-6}
\T\B & 0.08 & 0.6621 & 0.0028 & 4-10 &  0.14\\
\cline{2-6}
\T\B & 0.06 & 0.5759 & 0.0029 & 4-10 &  0.14\\
\cline{2-6}
\T\B & 0.05 & 0.5279 & 0.0029 & 4-10 &  0.13\\
\cline{2-6}
\T\B & 0.04 & 0.4750 & 0.0030 & 4-10 &  0.12\\
\cline{2-6}
\T\B & 0.03 & 0.4151 & 0.0032 & 4-10 &  0.12\\
\cline{2-6}
\T\B & 0.02 & 0.3420 & 0.0039 & 4-10 &  0.13\\
\hline
\T\B 1,6,9,10 & 0.20 & 1.0536 & 0.0029 & 3-9 & 0.10 \\
\cline{2-6}
\T\B & 0.16 & 0.9380 & 0.0029 & 3-9 &  0.12\\
\cline{2-6}
\T\B & 0.12 & 0.8100 & 0.0030 & 3-9 &  0.17\\
\cline{2-6}
\T\B & 0.10 & 0.7392 & 0.0031 & 3-9 &  0.20\\
\cline{2-6}
\T\B & 0.08 & 0.6621 & 0.0032 & 3-9 &  0.25\\
\cline{2-6}
\T\B & 0.06 & 0.5757 & 0.0033 & 3-9 &  0.29\\
\cline{2-6}
\T\B & 0.05 & 0.5274 & 0.0034 & 3-9 &  0.32\\
\cline{2-6}
\T\B & 0.04 & 0.4742 & 0.0036 & 3-9 &  0.35\\
\cline{2-6}
\T\B & 0.03 & 0.4136 & 0.0039 & 3-9 &  0.40\\
\cline{2-6}
\T\B & 0.02 & 0.3414 & 0.0052 & 3-9 &  0.49\\
\hline
\T\B 1,4,6,9,12 & 0.20 & 1.0534 & 0.0023 & 3-9 &  0.10\\
\cline{2-6}
\T\B & 0.16 & 0.9377 & 0.0024 & 3-9 &  0.13\\
\cline{2-6}
\T\B & 0.12 & 0.8096 & 0.0024 & 3-9 &  0.17\\
\cline{2-6}
\T\B & 0.10 & 0.7390 & 0.0028 & 3-9 &  0.20\\
\cline{2-6}
\T\B & 0.08 & 0.6619 & 0.0028 & 3-9 &  0.24\\
\cline{2-6}
\T\B & 0.06 & 0.5755 & 0.0029 & 3-9 &  0.28\\
\cline{2-6}
\T\B & 0.05 & 0.5271 & 0.0030 & 3-9 &  0.30\\
\cline{2-6}
\T\B & 0.04 & 0.4744 & 0.0036 & 3-9 &  0.33\\
\cline{2-6}
\T\B & 0.03 & 0.4140 & 0.0039 & 3-9 &  0.38\\
\cline{2-6}
\T\B & 0.02 & 0.3418 & 0.0050 & 3-9 &  0.48\\
\hline
\end{tabular}
\end{ruledtabular}
\end{center}
\caption{
Groundstate in the $0^{-+}$ channel (pion).
\label{ps_groundstate}}
\end{table}

\begin{table}[tbh]
\begin{center}
\begin{ruledtabular}
\begin{tabular}{c|c|c|c|c|c}
interpolators & $m_q$ & mass [GeV] & error & fitrange & $\chi^2/\mathrm{d.o.f.}$ \\
\hline\hline
\T\B 1,4,5,6 & 0.20 & 1.974 & 0.062 & 3-7 & 0.10\\
\cline{2-6}
\T\B & 0.16 & 1.891 & 0.069 & 3-7 & 0.09 \\
\cline{2-6}
\T\B & 0.12 & 1.813 & 0.079 & 3-7 & 0.04 \\
\cline{2-6}
\T\B & 0.10 & 1.778 & 0.086 & 3-7 & 0.02\\
\cline{2-6}
\T\B & 0.08 & 1.743 & 0.088 & 3-6 & 0.01\\
\cline{2-6}
\T\B & 0.06 & 1.698 & 0.100 & 3-6 & 0.02\\
\hline
\T\B 1,6,9,10 & 0.20 & 1.999 & 0.047 & 3-7 & 0.02\\
\cline{2-6}
\T\B & 0.16 & 1.917 & 0.054 & 3-7 & 0.01 \\
\cline{2-6}
\T\B & 0.12 & 1.834 & 0.062 & 3-6 & 0.07 \\
\cline{2-6}
\T\B & 0.10 & 1.791 & 0.070 & 3-6 & 0.14\\
\cline{2-6}
\T\B & 0.08 & 1.744 & 0.082 & 3-6 & 0.21\\
\cline{2-6}
\T\B & 0.06 & 1.675 & 0.101 & 3-6 & 0.15\\
\cline{2-6}
\T\B & 0.05 & 1.618 & 0.116 & 3-6 & 0.05 \\
\hline
\T\B 1,4,6,9,12 & 0.20 & 1.972 & 0.064 & 3-7 & 0.11\\
\cline{2-6}
\T\B & 0.16 & 1.888 & 0.069 & 3-7 & 0.10\\
\cline{2-6}
\T\B & 0.12 & 1.805 & 0.078 & 3-7 & 0.03\\
\cline{2-6}
\T\B & 0.10 & 1.766 & 0.084 & 3-7 & 0.05\\
\cline{2-6}
\T\B & 0.08 & 1.730 & 0.088 & 3-6 & 0.06\\
\cline{2-6}
\T\B & 0.06 & 1.684 & 0.104& 3-6 & 0.03\\
\cline{2-6}
\T\B & 0.05 & 1.646 & 0.117 & 3-6 &  0.07\\
\cline{2-6}
\T\B & 0.04 & 1.542 & 0.133 & 3-6 & 0.04 \\
\hline
\end{tabular}
\end{ruledtabular}
\end{center}
\caption{
First excited state in the $0^{-+}$ channel.
\label{ps_first}}
\end{table}

\begin{table}[tbh]
\begin{center}
\begin{ruledtabular}
\begin{tabular}{c|c|c|c|c|c}
interpolators & $m_q$ & mass [GeV] & error & fitrange & $\chi^2/\mathrm{d.o.f.}$ \\
\hline\hline
\T\B 1,6,9,10 & 0.20 & 2.266 & 0.108 & 3-7 & 0.01\\
\cline{2-6}
\T\B & 0.16 & 2.199 & 0.109 & 3-7 &  0.02\\
\cline{2-6}
\T\B & 0.12 & 2.142 & 0.120 & 3-6 & 0.07 \\
\cline{2-6}
\T\B & 0.10 & 2.113 & 0.134 & 3-6 & 0.12\\
\cline{2-6}
\T\B & 0.08 & 2.082 & 0.162  & 3-6 & 0.14\\
\hline
\T\B 1,4,6,9,12 & 0.20 & 2.291 & 0.089 & 3-6 & 0.08\\
\cline{2-6}
\T\B & 0.16 & 2.221 & 0.096 & 3-6 & 0.08\\
\cline{2-6}
\T\B & 0.12 & 2.155 & 0.111 & 3-6 & 0.06\\
\cline{2-6}
\T\B & 0.10 & 2.125 & 0.126 & 3-6 & 0.04\\
\cline{2-6}
\T\B & 0.08 & 2.098 & 0.154 & 3-6 & 0.01\\
\hline
\end{tabular}
\end{ruledtabular}
\end{center}
\caption{
Second excited state in the $0^{-+}$ channel.
\label{ps_second}}
\end{table}

\begin{table}[tbh]
\begin{center}
\begin{ruledtabular}
\begin{tabular}{c|c|c|c|c|c}
interpolators & $m_q$ & mass [GeV] & error & fitrange & $\chi^2/\mathrm{d.o.f.}$ \\
\hline\hline
\T\B 8,10,11 & 0.20 & 1.676 & 0.017 & 4-7 & 0.07\\
\cline{2-6}
\T\B & 0.16 & 1.599 & 0.019 & 4-7 & 0.14\\
\cline{2-6}
\T\B & 0.12 & 1.524 & 0.024 & 4-7 &  0.28\\
\cline{2-6}
\T\B & 0.10 & 1.487 & 0.028 & 4-7 & 0.37\\
\cline{2-6}
\T\B & 0.08 & 1.451 & 0.036 & 4-7 & 0.41\\
\cline{2-6}
\T\B & 0.06 & 1.416 & 0.051 & 4-7 & 0.34\\
\cline{2-6}
\T\B & 0.05 & 1.393 & 0.064 & 4-7 & 0.24 \\
\cline{2-6}
\T\B & 0.04 & 1.357 & 0.081 & 4-7 & 0.15 \\
\hline
\end{tabular}
\end{ruledtabular}
\end{center}
\caption{
Ground state in the $0^{++}$ channel.
\label{scalar}}
\end{table}

\begin{table}[tbh]
\begin{center}
\begin{ruledtabular}
\begin{tabular}{c|c|c|c|c|c}
interpolators & $m_q$ & mass [GeV] & error & fitrange & $\chi^2/\mathrm{d.o.f.}$ \\
\hline\hline
\T\B 1,4,5,6 & 0.20 & 1.251 & 0.0041 & 3-9 & 0.01\\
\cline{2-6}
\T\B & 0.16 & 1.165 & 0.004 & 3-9 & 0.01 \\
\cline{2-6}
\T\B & 0.12 & 1.079 & 0.005 & 3-9 &  0.01\\
\cline{2-6}
\T\B & 0.10 & 1.036 & 0.005 & 3-9 & 0.01\\
\cline{2-6}
\T\B & 0.08 & 0.993 & 0.006 & 3-9 & 0.01\\
\cline{2-6}
\T\B & 0.06 & 0.950 & 0.007 & 3-9 & 0.02\\
\cline{2-6}
\T\B & 0.05 & 0.931 & 0.009 & 3-6 & 0.01 \\
\cline{2-6}
\T\B & 0.04 & 0.912 & 0.010 & 3-6 &  0.01\\
\cline{2-6}
\T\B & 0.03 & 0.893 & 0.012 & 3-6 & 0.02 \\
\cline{2-6}
\T\B & 0.02 & 0.871 & 0.016 & 3-6 & 0.06\\
\hline
\T\B 1,2,7,8,11 & 0.20 & 1.251 & 0.003 & 3-12 & 0.02\\
\cline{2-6}
\T\B & 0.16 & 1.165 & 0.004 & 3-12 & 0.04 \\
\cline{2-6}
\T\B & 0.12 & 1.079 & 0.005 & 3-12 & 0.06 \\
\cline{2-6}
\T\B & 0.10 & 1.036 & 0.005 & 3-12 & 0.07 \\
\cline{2-6}
\T\B & 0.08 & 0.993 & 0.006 & 3-12 & 0.07 \\
\cline{2-6}
\T\B & 0.06 & 0.952 & 0.008 & 3-12 & 0.06 \\
\cline{2-6}
\T\B & 0.05 & 0.933 & 0.009 & 3-12 &  0.04 \\
\cline{2-6}
\T\B & 0.04 & 0.914 & 0.011 & 3-12 & 0.03 \\
\cline{2-6}
\T\B & 0.03 & 0.896 & 0.013 & 3-12 &  0.02 \\
\cline{2-6}
\T\B & 0.02 & 0.881 & 0.017 & 3-12 &  0.07 \\
\hline
\end{tabular}
\end{ruledtabular}
\end{center}
\caption{
Ground state in the $1^{--}$ channel.
\label{vector_ground}}
\end{table}

\begin{table}[tbh]
\begin{center}
\begin{ruledtabular}
\begin{tabular}{c|c|c|c|c|c}
interpolators & $m_q$ & mass [GeV] & error & fitrange & $\chi^2/\mathrm{d.o.f.}$ \\
\hline\hline
\T\B 1,4,5,6 & 0.20 & 2.094 & 0.053 & 3-6 & 0.05\\
\cline{2-6}
\T\B & 0.16 & 2.033 & 0.059 & 3-6 & 0.06 \\
\cline{2-6}
\T\B & 0.12 & 1.978 & 0.065 & 3-6 & 0.06 \\
\cline{2-6}
\T\B & 0.10 & 1.954 & 0.069 & 3-6 & 0.05\\
\cline{2-6}
\T\B & 0.08 & 1.935 & 0.074 & 3-6 & 0.04\\
\hline
\T\B 1,2,7,8,11 & 0.20 & 2.091 & 0.037 & 3-7 & 0.05\\
\cline{2-6}
\T\B & 0.16 & 2.033 & 0.040 & 3-7 &  0.06 \\
\cline{2-6}
\T\B & 0.12 & 1.985 & 0.044 & 3-7 & 0.11 \\
\cline{2-6}
\T\B & 0.10 & 1.968 & 0.046 & 3-7 & 0.16 \\
\cline{2-6}
\T\B & 0.08 & 1.957 & 0.049 & 3-7 & 0.21 \\
\cline{2-6}
\T\B & 0.06 & 1.941 & 0.055 & 3-6 & 0.14 \\
\hline
\end{tabular}
\end{ruledtabular}
\end{center}
\caption{
First excited state in the $1^{--}$ channel.
\label{vector_first}}
\end{table}

\begin{table}[tbh]
\begin{center}
\begin{ruledtabular}
\begin{tabular}{c|c|c|c|c|c}
interpolators & $m_q$ & mass [GeV] & error & fitrange & $\chi^2/\mathrm{d.o.f.}$ \\
\hline\hline
\T\B 1,4,5,6 & 0.20 & 2.209 & 0.065 & 3-6 & 1.49\\
\cline{2-6}
\T\B & 0.16 & 2.117 & 0.063 & 3-6 & 1.50 \\
\cline{2-6}
\T\B & 0.12 & 2.036 & 0.071 &  3-6 & 0.98 \\
\cline{2-6}
\T\B & 0.10 & 1.993 & 0.079 & 3-6 & 0.62\\
\cline{2-6}
\T\B & 0.08 & 1.948 & 0.093 & 3-6 & 0.25\\
\cline{2-6}
\T\B & 0.06 & 1.909 & 0.120 & 3-6 & 0.08\\
\hline
\T\B 1,2,7,8,11 & 0.20 & 2.178 & 0.029 & 3-7 & 1.25\\
\cline{2-6}
\T\B & 0.16 & 2.111 & 0.030 & 3-7 & 1.18 \\
\cline{2-6}
\T\B & 0.12 & 2.031 & 0.031 & 3-7 & 0.42 \\
\cline{2-6}
\T\B & 0.10 & 2.000 & 0.034 & 3-7 & 0.33 \\
\cline{2-6}
\T\B & 0.08 & 1.974 & 0.038 & 3-7 & 0.24 \\
\cline{2-6}
\T\B & 0.06 & 1.956 & 0.046 & 3-6 & 0.21 \\
\hline
\end{tabular}
\end{ruledtabular}
\end{center}
\caption{
Second excited state in the $1^{--}$ channel.
\label{vector second}}
\end{table}

\begin{table}[tbh]
\begin{center}
\begin{ruledtabular}
\begin{tabular}{c|c|c|c|c|c}
interpolators & $m_q$ & mass [GeV] & error & fitrange & $\chi^2/\mathrm{d.o.f.}$ \\
\hline\hline
\T\B 1,2,3 & 0.20 & 1.729 & 0.021 & 3-7 & 0.55\\
\cline{2-6}
\T\B & 0.16 & 1.652 & 0.023 & 3-7 & 0.47\\
\cline{2-6}
\T\B & 0.12 & 1.576 & 0.027 & 3-7 & 0.36\\
\cline{2-6}
\T\B & 0.10 & 1.539 & 0.031 & 3-7 & 0.30\\
\cline{2-6}
\T\B & 0.08 & 1.505 & 0.036 & 3-7 & 0.23\\
\cline{2-6}
\T\B & 0.06 & 1.478 & 0.041 & 3-6 & 0.26\\
\cline{2-6}
\T\B & 0.05 & 1.468 & 0.048 & 3-6 & 0.20\\
\cline{2-6}
\T\B & 0.04 & 1.461 & 0.058 & 3-6 & 0.13\\
\cline{2-6}
\T\B & 0.03 & 1.459 & 0.073 & 3-6 & 0.06\\
\cline{2-6}
\T\B & 0.02 & 1.469 & 0.099 & 3-6 & 0.02\\
\hline
\T\B 11 & 0.20 & 1.771 & 0.013 & 4-9 & 0.19\\
\cline{2-6}
\T\B & 0.16 & 1.700 & 0.015 & 4-9 & 0.26 \\
\cline{2-6}
\T\B & 0.12 & 1.629 & 0.018 & 4-9 & 0.34 \\
\cline{2-6}
\T\B & 0.10 & 1.592 & 0.020 & 4-9 & 0.34\\
\cline{2-6}
\T\B & 0.08 & 1.554 & 0.023 & 4-9 & 0.29\\
\cline{2-6}
\T\B & 0.06 & 1.512 & 0.028 & 4-9 & 0.16\\
\cline{2-6}
\T\B & 0.05 & 1.488 & 0.031 & 4-9 &  0.08\\
\cline{2-6}
\T\B & 0.04 & 1.462  & 0.036 & 4-9 & 0.04 \\
\cline{2-6}
\T\B & 0.03 & 1.432 & 0.043 & 4-9 &  0.16\\
\cline{2-6}
\T\B & 0.02 & 1.390 & 0.056 & 4-9 &  0.76\\
\hline
\end{tabular}
\end{ruledtabular}
\end{center}
\caption{
Ground state in the $1^{++}$ channel.
\label{axial_ground}}
\end{table}

\begin{table}[tbh]
\begin{center}
\begin{ruledtabular}
\begin{tabular}{c|c|c|c|c|c}
interpolators & $m_q$ & mass [GeV] & error & fitrange & $\chi^2/\mathrm{d.o.f.}$ \\
\hline\hline
\T\B 1,2,3 & 0.20 & 2.322 & 0.088 & 3-6 & 0.06\\
\cline{2-6}
\T\B & 0.16 & 2.265 & 0.103 & 3-6 & 0.04\\
\cline{2-6}
\T\B & 0.12 & 2.236 & 0.123 & 3-6 & 0.02\\
\cline{2-6}
\T\B & 0.10 & 2.241 & 0.136 & 3-6 & 0.03\\
\cline{2-6}
\T\B & 0.08 & 2.270 & 0.149 & 3-6 & 0.09\\
\hline
\T\B 1,2,11 & 0.20 & 2.342 & 0.079 & 3-6 & 0.01\\
\cline{2-6}
\T\B & 0.16 & 2.288 & 0.090 & 3-6 & 0.02\\
\cline{2-6}
\T\B & 0.12 & 2.250 & 0.111 & 3-6 & 0.04\\
\cline{2-6}
\T\B & 0.10 & 2.247 & 0.130 & 3-6 & 0.07\\
\hline
\end{tabular}
\end{ruledtabular}
\end{center}
\caption{
First excited state in the $1^{++}$ channel.
\label{axial_first}}
\end{table}

\bibliography{revtex_draft}

\begin{thebibliography}{45}
\expandafter\ifx\csname natexlab\endcsname\relax\def\natexlab#1{#1}\fi
\expandafter\ifx\csname bibnamefont\endcsname\relax
  \def\bibnamefont#1{#1}\fi
\expandafter\ifx\csname bibfnamefont\endcsname\relax
  \def\bibfnamefont#1{#1}\fi
\expandafter\ifx\csname citenamefont\endcsname\relax
  \def\citenamefont#1{#1}\fi
\expandafter\ifx\csname url\endcsname\relax
  \def\url#1{\texttt{#1}}\fi
\expandafter\ifx\csname urlprefix\endcsname\relax\def\urlprefix{URL }\fi
\providecommand{\bibinfo}[2]{#2}
\providecommand{\eprint}[2][]{\url{#2}}

\bibitem[{\citenamefont{Asakawa et~al.}(2001)\citenamefont{Asakawa, Hatsuda,
  and Nakahara}}]{Asakawa:2000tr}
\bibinfo{author}{\bibfnamefont{M.}~\bibnamefont{Asakawa}},
  \bibinfo{author}{\bibfnamefont{T.}~\bibnamefont{Hatsuda}}, \bibnamefont{and}
  \bibinfo{author}{\bibfnamefont{Y.}~\bibnamefont{Nakahara}},
  \bibinfo{journal}{Prog. Part. Nucl. Phys.} \textbf{\bibinfo{volume}{46}},
  \bibinfo{pages}{459} (\bibinfo{year}{2001}), \eprint{hep-lat/0011040}.

\bibitem[{\citenamefont{Lepage et~al.}(2002)}]{Lepage:2001ym}
\bibinfo{author}{\bibfnamefont{G.}~\bibnamefont{Lepage}} \bibnamefont{et~al.},
  \bibinfo{journal}{Nucl. Phys. Proc. Suppl.} \textbf{\bibinfo{volume}{106}},
  \bibinfo{pages}{12} (\bibinfo{year}{2002}), \eprint{hep-lat/0110175}.

\bibitem[{\citenamefont{Chen et~al.}(2004)}]{Chen:2004gp}
\bibinfo{author}{\bibfnamefont{Y.}~\bibnamefont{Chen}} \bibnamefont{et~al.}
  (\bibinfo{year}{2004}), \eprint{hep-lat/0405001}.

\bibitem[{\citenamefont{Fleming}(2004)}]{Fleming:2004hs}
\bibinfo{author}{\bibfnamefont{G.~T.} \bibnamefont{Fleming}}
  (\bibinfo{year}{2004}), \eprint{hep-lat/0403023}.

\bibitem[{\citenamefont{Lin and Cohen}(2007)}]{Lin:2007iq}
\bibinfo{author}{\bibfnamefont{H.-W.} \bibnamefont{Lin}} \bibnamefont{and}
  \bibinfo{author}{\bibfnamefont{S.~D.} \bibnamefont{Cohen}}
  (\bibinfo{year}{2007}), \eprint{arXiv:0709.1902 [hep-lat]}.

\bibitem[{\citenamefont{von Hippel et~al.}(2007)\citenamefont{von Hippel,
  Lewis, and Petry}}]{vonHippel:2007dz}
\bibinfo{author}{\bibfnamefont{G.~M.} \bibnamefont{von Hippel}},
  \bibinfo{author}{\bibfnamefont{R.}~\bibnamefont{Lewis}}, \bibnamefont{and}
  \bibinfo{author}{\bibfnamefont{R.~G.} \bibnamefont{Petry}},
  \bibinfo{journal}{PoS(LATTICE2007)043}  (\bibinfo{year}{2007}),
  \eprint{arXiv:0710.0014 [hep-lat]}.

\bibitem[{\citenamefont{Michael}(1985)}]{Michael:1985ne}
\bibinfo{author}{\bibfnamefont{C.}~\bibnamefont{Michael}},
  \bibinfo{journal}{Nucl. Phys.} \textbf{\bibinfo{volume}{B259}},
  \bibinfo{pages}{58} (\bibinfo{year}{1985}).

\bibitem[{\citenamefont{L{\"u}scher and Wolff}(1990)}]{Luscher:1990ck}
\bibinfo{author}{\bibfnamefont{M.}~\bibnamefont{L{\"u}scher}} \bibnamefont{and}
  \bibinfo{author}{\bibfnamefont{U.}~\bibnamefont{Wolff}},
  \bibinfo{journal}{Nucl. Phys.} \textbf{\bibinfo{volume}{B339}},
  \bibinfo{pages}{222} (\bibinfo{year}{1990}).

\bibitem[{\citenamefont{Burch et~al.}(2006{\natexlab{a}})}]{Burch:2006dg}
\bibinfo{author}{\bibfnamefont{T.}~\bibnamefont{Burch}} \bibnamefont{et~al.},
  \bibinfo{journal}{Phys. Rev.} \textbf{\bibinfo{volume}{D73}},
  \bibinfo{pages}{094505} (\bibinfo{year}{2006}{\natexlab{a}}),
  \eprint{hep-lat/0601026}.

\bibitem[{\citenamefont{Burch et~al.}(2006{\natexlab{b}})}]{Burch:2006cc}
\bibinfo{author}{\bibfnamefont{T.}~\bibnamefont{Burch}} \bibnamefont{et~al.},
  \bibinfo{journal}{Phys. Rev.} \textbf{\bibinfo{volume}{D74}},
  \bibinfo{pages}{014504} (\bibinfo{year}{2006}{\natexlab{b}}),
  \eprint{hep-lat/0604019}.

\bibitem[{\citenamefont{Burch et~al.}(2005)}]{Burch:2005vn}
\bibinfo{author}{\bibfnamefont{T.}~\bibnamefont{Burch}} \bibnamefont{et~al.},
  \bibinfo{journal}{Nucl. Phys.} \textbf{\bibinfo{volume}{A755}},
  \bibinfo{pages}{481} (\bibinfo{year}{2005}), \eprint{nucl-th/0501025}.

\bibitem[{\citenamefont{Burch et~al.}(2004)}]{Burch:2004he}
\bibinfo{author}{\bibfnamefont{T.}~\bibnamefont{Burch}} \bibnamefont{et~al.}
  (\bibinfo{collaboration}{Bern-Graz-Regensburg}), \bibinfo{journal}{Phys.
  Rev.} \textbf{\bibinfo{volume}{D70}}, \bibinfo{pages}{054502}
  (\bibinfo{year}{2004}), \eprint{hep-lat/0405006}.

\bibitem[{\citenamefont{Liao and Manke}(2002)}]{Liao:2002rj}
\bibinfo{author}{\bibfnamefont{X.}~\bibnamefont{Liao}} \bibnamefont{and}
  \bibinfo{author}{\bibfnamefont{T.}~\bibnamefont{Manke}}
  (\bibinfo{year}{2002}), \eprint{hep-lat/0210030}.

\bibitem[{\citenamefont{Dudek et~al.}(2008)\citenamefont{Dudek, Edwards,
  Mathur, and Richards}}]{Dudek:2007wv}
\bibinfo{author}{\bibfnamefont{J.~J.} \bibnamefont{Dudek}},
  \bibinfo{author}{\bibfnamefont{R.~G.} \bibnamefont{Edwards}},
  \bibinfo{author}{\bibfnamefont{N.}~\bibnamefont{Mathur}}, \bibnamefont{and}
  \bibinfo{author}{\bibfnamefont{D.~G.} \bibnamefont{Richards}},
  \bibinfo{journal}{Phys. Rev.} \textbf{\bibinfo{volume}{D77}},
  \bibinfo{pages}{034501} (\bibinfo{year}{2008}), \eprint{arXiv:0707.4162
  [hep-lat]}.

\bibitem[{\citenamefont{Wingate et~al.}(1995)\citenamefont{Wingate, DeGrand,
  Collins, and Heller}}]{Wingate:1995hy}
\bibinfo{author}{\bibfnamefont{M.}~\bibnamefont{Wingate}},
  \bibinfo{author}{\bibfnamefont{T.~A.} \bibnamefont{DeGrand}},
  \bibinfo{author}{\bibfnamefont{S.}~\bibnamefont{Collins}}, \bibnamefont{and}
  \bibinfo{author}{\bibfnamefont{U.~M.} \bibnamefont{Heller}},
  \bibinfo{journal}{Phys. Rev. Lett.} \textbf{\bibinfo{volume}{74}},
  \bibinfo{pages}{4596} (\bibinfo{year}{1995}), \eprint{hep-ph/9502274}.

\bibitem[{\citenamefont{Burch et~al.}(2006{\natexlab{c}})\citenamefont{Burch,
  Hagen, and Sch{\"{a}}fer}}]{Burch:2006rb}
\bibinfo{author}{\bibfnamefont{T.}~\bibnamefont{Burch}},
  \bibinfo{author}{\bibfnamefont{C.}~\bibnamefont{Hagen}}, \bibnamefont{and}
  \bibinfo{author}{\bibfnamefont{A.}~\bibnamefont{Sch{\"{a}}fer}},
  \bibinfo{journal}{PoS} \textbf{\bibinfo{volume}{LAT2006}},
  \bibinfo{pages}{177} (\bibinfo{year}{2006}{\natexlab{c}}),
  \eprint{hep-lat/0609014}.

\bibitem[{\citenamefont{Burch et~al.}(2007)\citenamefont{Burch, Ehmann, Hagen,
  Hetzenegger, and Sch{\"{a}}fer}}]{Burch:2007dh}
\bibinfo{author}{\bibfnamefont{T.}~\bibnamefont{Burch}},
  \bibinfo{author}{\bibfnamefont{C.}~\bibnamefont{Ehmann}},
  \bibinfo{author}{\bibfnamefont{C.}~\bibnamefont{Hagen}},
  \bibinfo{author}{\bibfnamefont{M.}~\bibnamefont{Hetzenegger}},
  \bibnamefont{and}
  \bibinfo{author}{\bibfnamefont{A.}~\bibnamefont{Sch{\"{a}}fer}},
  \bibinfo{journal}{PoS(LATTICE2007)103}  (\bibinfo{year}{2007}),
  \eprint{arXiv:0709.0664 [hep-lat]}.

\bibitem[{\citenamefont{Lacock et~al.}(1996)\citenamefont{Lacock, Michael,
  Boyle, and Rowland}}]{Lacock:1996vy}
\bibinfo{author}{\bibfnamefont{P.}~\bibnamefont{Lacock}},
  \bibinfo{author}{\bibfnamefont{C.}~\bibnamefont{Michael}},
  \bibinfo{author}{\bibfnamefont{P.}~\bibnamefont{Boyle}}, \bibnamefont{and}
  \bibinfo{author}{\bibfnamefont{P.}~\bibnamefont{Rowland}}
  (\bibinfo{collaboration}{UKQCD}), \bibinfo{journal}{Phys. Rev.}
  \textbf{\bibinfo{volume}{D54}}, \bibinfo{pages}{6997} (\bibinfo{year}{1996}),
  \eprint{hep-lat/9605025}.

\bibitem[{\citenamefont{Basak et~al.}(2005{\natexlab{a}})}]{Basak:2005ir}
\bibinfo{author}{\bibfnamefont{S.}~\bibnamefont{Basak}} \bibnamefont{et~al.}
  (\bibinfo{collaboration}{Lattice Hadron Physics (LHPC)}),
  \bibinfo{journal}{Phys. Rev.} \textbf{\bibinfo{volume}{D72}},
  \bibinfo{pages}{074501} (\bibinfo{year}{2005}{\natexlab{a}}),
  \eprint{hep-lat/0508018}.

\bibitem[{\citenamefont{Basak et~al.}(2005{\natexlab{b}})}]{Basak:2005aq}
\bibinfo{author}{\bibfnamefont{S.}~\bibnamefont{Basak}} \bibnamefont{et~al.},
  \bibinfo{journal}{Phys. Rev.} \textbf{\bibinfo{volume}{D72}},
  \bibinfo{pages}{094506} (\bibinfo{year}{2005}{\natexlab{b}}),
  \eprint{hep-lat/0506029}.

\bibitem[{\citenamefont{Basak et~al.}(2007)}]{Basak:2007kj}
\bibinfo{author}{\bibfnamefont{S.}~\bibnamefont{Basak}} \bibnamefont{et~al.},
  \bibinfo{journal}{Phys. Rev.} \textbf{\bibinfo{volume}{D76}},
  \bibinfo{pages}{074504} (\bibinfo{year}{2007}), \eprint{arXiv:0709.0008
  [hep-lat]}.

\bibitem[{\citenamefont{Basak et~al.}(2006)}]{Basak:2006ww}
\bibinfo{author}{\bibfnamefont{S.}~\bibnamefont{Basak}} \bibnamefont{et~al.}
  (\bibinfo{year}{2006}), \eprint{hep-lat/0609052}.

\bibitem[{\citenamefont{Gattringer et~al.}(2007)\citenamefont{Gattringer,
  Glozman, Lang, Mohler, and Prelovsek}}]{Gattringer:2007td}
\bibinfo{author}{\bibfnamefont{C.}~\bibnamefont{Gattringer}},
  \bibinfo{author}{\bibfnamefont{L.~Y.} \bibnamefont{Glozman}},
  \bibinfo{author}{\bibfnamefont{C.~B.} \bibnamefont{Lang}},
  \bibinfo{author}{\bibfnamefont{D.}~\bibnamefont{Mohler}}, \bibnamefont{and}
  \bibinfo{author}{\bibfnamefont{S.}~\bibnamefont{Prelovsek}},
  \bibinfo{journal}{PoS(LATTICE2007)123}  (\bibinfo{year}{2007}),
  \eprint{arXiv:0709.4456 [hep-lat]}.

\bibitem[{\citenamefont{Gattringer}(2007)}]{Gattringer:2007da}
\bibinfo{author}{\bibfnamefont{C.}~\bibnamefont{Gattringer}},
  \bibinfo{journal}{EPJA (to be published)}  (\bibinfo{year}{2007}),
  \eprint{arXiv:0711.0622 [hep-lat]}.

\bibitem[{\citenamefont{Lang}(2007)}]{Lang:2007mq}
\bibinfo{author}{\bibfnamefont{C.~B.} \bibnamefont{Lang}},
  \bibinfo{journal}{Prog. Part. Nucl. Phys. (to be published)}
  (\bibinfo{year}{2007}), \eprint{arXiv:0711.3091 [nucl-th]}.

\bibitem[{\citenamefont{Bardeen et~al.}(2001)\citenamefont{Bardeen, Duncan,
  Eichten, Isgur, and Thacker}}]{Bardeen:2001jm}
\bibinfo{author}{\bibfnamefont{W.~A.} \bibnamefont{Bardeen}},
  \bibinfo{author}{\bibfnamefont{A.}~\bibnamefont{Duncan}},
  \bibinfo{author}{\bibfnamefont{E.}~\bibnamefont{Eichten}},
  \bibinfo{author}{\bibfnamefont{N.}~\bibnamefont{Isgur}}, \bibnamefont{and}
  \bibinfo{author}{\bibfnamefont{H.}~\bibnamefont{Thacker}},
  \bibinfo{journal}{Phys. Rev.} \textbf{\bibinfo{volume}{D65}},
  \bibinfo{pages}{014509} (\bibinfo{year}{2001}), \eprint{hep-lat/0106008}.

\bibitem[{\citenamefont{Burch et~al.}(2006{\natexlab{d}})\citenamefont{Burch,
  Gattringer, Glozman, Hagen, and Lang}}]{Burch:2005wd}
\bibinfo{author}{\bibfnamefont{T.}~\bibnamefont{Burch}},
  \bibinfo{author}{\bibfnamefont{C.}~\bibnamefont{Gattringer}},
  \bibinfo{author}{\bibfnamefont{L.~Y.} \bibnamefont{Glozman}},
  \bibinfo{author}{\bibfnamefont{C.}~\bibnamefont{Hagen}}, \bibnamefont{and}
  \bibinfo{author}{\bibfnamefont{C.~B.} \bibnamefont{Lang}},
  \bibinfo{journal}{Phys. Rev.} \textbf{\bibinfo{volume}{D73}},
  \bibinfo{pages}{017502} (\bibinfo{year}{2006}{\natexlab{d}}),
  \eprint{hep-lat/0511054}.

\bibitem[{\citenamefont{G{\"u}sken et~al.}(1989)}]{Gusken:1989ad}
\bibinfo{author}{\bibfnamefont{S.}~\bibnamefont{G{\"u}sken}}
  \bibnamefont{et~al.}, \bibinfo{journal}{Phys. Lett.}
  \textbf{\bibinfo{volume}{B227}}, \bibinfo{pages}{266} (\bibinfo{year}{1989}).

\bibitem[{\citenamefont{Best et~al.}(1997)}]{Best:1997qp}
\bibinfo{author}{\bibfnamefont{C.}~\bibnamefont{Best}} \bibnamefont{et~al.},
  \bibinfo{journal}{Phys. Rev.} \textbf{\bibinfo{volume}{D56}},
  \bibinfo{pages}{2743} (\bibinfo{year}{1997}), \eprint{hep-lat/9703014}.

\bibitem[{\citenamefont{Glozman}(2004)}]{Glozman:2003bt}
\bibinfo{author}{\bibfnamefont{L.~Y.} \bibnamefont{Glozman}},
  \bibinfo{journal}{Phys. Lett.} \textbf{\bibinfo{volume}{B587}},
  \bibinfo{pages}{69} (\bibinfo{year}{2004}), \eprint{hep-ph/0312354}.

\bibitem[{\citenamefont{Glozman}(2007)}]{Glozman:2007ek}
\bibinfo{author}{\bibfnamefont{L.~Y.} \bibnamefont{Glozman}},
  \bibinfo{journal}{Phys. Rept.} \textbf{\bibinfo{volume}{444}},
  \bibinfo{pages}{1} (\bibinfo{year}{2007}), \eprint{hep-ph/0701081}.

\bibitem[{\citenamefont{Glozman and Nefediev}(2007)}]{Glozman:2007at}
\bibinfo{author}{\bibfnamefont{L.~Y.} \bibnamefont{Glozman}} \bibnamefont{and}
  \bibinfo{author}{\bibfnamefont{A.~V.} \bibnamefont{Nefediev}},
  \bibinfo{journal}{Phys. Rev.} \textbf{\bibinfo{volume}{D76}},
  \bibinfo{pages}{096004} (\bibinfo{year}{2007}), \eprint{arXiv:0704.2673
  [hep-ph]}.

\bibitem[{\citenamefont{L{\"u}scher and Weisz}(1985)}]{Luscher:1984xn}
\bibinfo{author}{\bibfnamefont{M.}~\bibnamefont{L{\"u}scher}} \bibnamefont{and}
  \bibinfo{author}{\bibfnamefont{P.}~\bibnamefont{Weisz}},
  \bibinfo{journal}{Commun. Math. Phys.} \textbf{\bibinfo{volume}{97}},
  \bibinfo{pages}{59} (\bibinfo{year}{1985}), \bibinfo{note}{errata
  \textbf{98}, 433 (1985)}.

\bibitem[{\citenamefont{Curci et~al.}(1983)\citenamefont{Curci, Menotti, and
  Paffuti}}]{Curci:1983an}
\bibinfo{author}{\bibfnamefont{G.}~\bibnamefont{Curci}},
  \bibinfo{author}{\bibfnamefont{P.}~\bibnamefont{Menotti}}, \bibnamefont{and}
  \bibinfo{author}{\bibfnamefont{G.}~\bibnamefont{Paffuti}},
  \bibinfo{journal}{Phys. Lett.} \textbf{\bibinfo{volume}{B130}},
  \bibinfo{pages}{205} (\bibinfo{year}{1983}), \bibinfo{note}{errata
  \textbf{B135}, 516 (1984)}.

\bibitem[{\citenamefont{Gattringer et~al.}(2002)\citenamefont{Gattringer,
  Hoffmann, and Schaefer}}]{Gattringer:2001jf}
\bibinfo{author}{\bibfnamefont{C.}~\bibnamefont{Gattringer}},
  \bibinfo{author}{\bibfnamefont{R.}~\bibnamefont{Hoffmann}}, \bibnamefont{and}
  \bibinfo{author}{\bibfnamefont{S.}~\bibnamefont{Schaefer}},
  \bibinfo{journal}{Phys. Rev.} \textbf{\bibinfo{volume}{D65}},
  \bibinfo{pages}{094503} (\bibinfo{year}{2002}), \eprint{hep-lat/0112024}.

\bibitem[{\citenamefont{Gattringer}(2001)}]{Gattringer:2000js}
\bibinfo{author}{\bibfnamefont{C.}~\bibnamefont{Gattringer}},
  \bibinfo{journal}{Phys. Rev.} \textbf{\bibinfo{volume}{D63}},
  \bibinfo{pages}{114501} (\bibinfo{year}{2001}), \eprint{hep-lat/0003005}.

\bibitem[{\citenamefont{Gattringer et~al.}(2001)\citenamefont{Gattringer, Hip,
  and Lang}}]{Gattringer:2000qu}
\bibinfo{author}{\bibfnamefont{C.}~\bibnamefont{Gattringer}},
  \bibinfo{author}{\bibfnamefont{I.}~\bibnamefont{Hip}}, \bibnamefont{and}
  \bibinfo{author}{\bibfnamefont{C.~B.} \bibnamefont{Lang}},
  \bibinfo{journal}{Nucl. Phys.} \textbf{\bibinfo{volume}{B597}},
  \bibinfo{pages}{451} (\bibinfo{year}{2001}), \eprint{hep-lat/0007042}.

\bibitem[{\citenamefont{Gattringer et~al.}(2004)}]{Gattringer:2003qx}
\bibinfo{author}{\bibfnamefont{C.}~\bibnamefont{Gattringer}}
  \bibnamefont{et~al.} (\bibinfo{collaboration}{BGR}), \bibinfo{journal}{Nucl.
  Phys.} \textbf{\bibinfo{volume}{B677}}, \bibinfo{pages}{3}
  (\bibinfo{year}{2004}), \eprint{hep-lat/0307013}.

\bibitem[{\citenamefont{Lichtl}(2007)}]{Lichtl:2007jc}
\bibinfo{author}{\bibfnamefont{A.~C.} \bibnamefont{Lichtl}},
  \bibinfo{journal}{PoS(LATTICE2007)118}  (\bibinfo{year}{2007}),
  \eprint{arXiv:0711.4072 [hep-lat]}.

\bibitem[{\citenamefont{Bardeen et~al.}(2004)\citenamefont{Bardeen, Eichten,
  and Thacker}}]{Bardeen:2003qz}
\bibinfo{author}{\bibfnamefont{W.~A.} \bibnamefont{Bardeen}},
  \bibinfo{author}{\bibfnamefont{E.}~\bibnamefont{Eichten}}, \bibnamefont{and}
  \bibinfo{author}{\bibfnamefont{H.}~\bibnamefont{Thacker}},
  \bibinfo{journal}{Phys. Rev.} \textbf{\bibinfo{volume}{D69}},
  \bibinfo{pages}{054502} (\bibinfo{year}{2004}), \eprint{hep-lat/0307023}.

\bibitem[{\citenamefont{Prelovsek and Orginos}(2003)}]{Prelovsek:2002qs}
\bibinfo{author}{\bibfnamefont{S.}~\bibnamefont{Prelovsek}} \bibnamefont{and}
  \bibinfo{author}{\bibfnamefont{K.}~\bibnamefont{Orginos}}
  (\bibinfo{collaboration}{RBC}), \bibinfo{journal}{Nucl. Phys. Proc. Suppl.}
  \textbf{\bibinfo{volume}{119}}, \bibinfo{pages}{822} (\bibinfo{year}{2003}),
  \eprint{hep-lat/0209132}.

\bibitem[{\citenamefont{Prelovsek et~al.}(2004)\citenamefont{Prelovsek, Dawson,
  Izubuchi, Orginos, and Soni}}]{Prelovsek:2004jp}
\bibinfo{author}{\bibfnamefont{S.}~\bibnamefont{Prelovsek}},
  \bibinfo{author}{\bibfnamefont{C.}~\bibnamefont{Dawson}},
  \bibinfo{author}{\bibfnamefont{T.}~\bibnamefont{Izubuchi}},
  \bibinfo{author}{\bibfnamefont{K.}~\bibnamefont{Orginos}}, \bibnamefont{and}
  \bibinfo{author}{\bibfnamefont{A.}~\bibnamefont{Soni}},
  \bibinfo{journal}{Phys. Rev.} \textbf{\bibinfo{volume}{D70}},
  \bibinfo{pages}{094503} (\bibinfo{year}{2004}), \eprint{hep-lat/0407037}.

\bibitem[{\citenamefont{Mathur et~al.}(2007)}]{Mathur:2006bs}
\bibinfo{author}{\bibfnamefont{N.}~\bibnamefont{Mathur}} \bibnamefont{et~al.},
  \bibinfo{journal}{Phys. Rev.} \textbf{\bibinfo{volume}{D76}},
  \bibinfo{pages}{114505} (\bibinfo{year}{2007}), \eprint{hep-ph/0607110}.

\bibitem[{\citenamefont{Frigori et~al.}(2007)}]{Frigori:2007wa}
\bibinfo{author}{\bibfnamefont{R.}~\bibnamefont{Frigori}} \bibnamefont{et~al.},
  \bibinfo{journal}{PoS(LATTICE2007)114}  (\bibinfo{year}{2007}),
  \eprint{arXiv:0709.4582 [hep-lat]}.

\bibitem[{\citenamefont{McNeile}(2007)}]{McNeile:2007fu}
\bibinfo{author}{\bibfnamefont{C.}~\bibnamefont{McNeile}},
  \bibinfo{journal}{PoS(LATTICE2007)019}  (\bibinfo{year}{2007}),
  \eprint{arXiv:0710.0985 [hep-lat]}.

\end{thebibliography}
\bibstyle{apsrev}

\end{document}